\RequirePackage{fix-cm}
\documentclass[twocolumn]{svjour3}          
\smartqed  
\usepackage{graphicx}
\usepackage[misc]{ifsym}
\usepackage{amssymb}
\usepackage{amsmath}
\usepackage{enumitem}
\usepackage{booktabs}
\usepackage{subcaption}
\usepackage{tcolorbox}
\usepackage{framed}

\begin{document}

\title{Strength Is Not Value: Rethinking Approximate Functional Dependency Measures for Error Detection
}


\author{Xiaolong Wan         \and
        Xixian Han 
}



\institute{Xiaolong Wan \at
	Harbin Institute of Technology, China. \\
	\email{wxl@hit.edu.cn}           
	\and
	Xixian Han \at
	Harbin Institute of Technology, China. \\
	\email{hanxx@hit.edu.cn}           
}


\date{Received: date / Accepted: date}

\maketitle

\begin{abstract}
Approximate functional dependency (AFD) strength measures are commonly used to prioritize candidate dependencies, including when selected dependencies support downstream tasks such as error detection. This practice implicitly assumes that dependencies ranked higher by strength are also better candidates for the task they are intended to support. We test this assumption directly for error detection. We evaluate detection value through precision and recall without imposing a single weighting on different error types, and introduce an explicit false-positive/false-negative cost only when a single candidate-selection objective is required. Across three real-world relations from an established AFD benchmark, $1{,}155$ corruption-and-detection runs, and $513{,}480$ candidate-level evaluations, we find a clear asymmetry. Strength measures tend to track precision positively, but their associations with recall are substantially less stable, with both magnitude and direction varying across measures and experimental settings. Strength-based prioritization is also unreliable at the ranking level: top-$k$ sets chosen by strength often have little overlap with those favored by actual detection performance, and strength-based selections can incur substantial regret. An important part of the recall mismatch is associated with structural detection opportunity under the candidate equivalence-class partition, which tracks recall more consistently overall than the strength measures we evaluate. Corruption mechanism and several static structural properties account for selected effects but do not provide a common explanation across settings, leaving part of the mismatch unresolved. Overall, AFD strength and detection value are distinct and should be evaluated separately.
	
\end{abstract}

\section{Introduction}
\label{sec:intro}

If an attribute set $X$ approximately determines another attribute $Y$ in a relation, a tuple whose $Y$-value departs from what $X$ typically implies is a natural candidate for an error. This observation underlies a substantial line of work on dependency-based data cleaning~\cite{DBLP:journals/tkde/CaruccioDP16,DBLP:books/acm/IlyasC19,DBLP:conf/sigmod/ZhangGR20}, where approximate functional dependencies (AFDs) provide detection rules without labeled examples or an external clean reference. The relation itself is the only resource such rules need. Putting this idea into practice, however, requires distinguishing which candidate dependencies are worth using. Not every dependency admitted by a finite relation reflects a genuine regularity, and some may hold only by coincidence~\cite{DBLP:conf/kdd/MandrosBV17}. Even a relation of modest width admits a combinatorially large candidate space~\cite{DBLP:journals/cj/HuhtalaKPT99}. Candidate dependencies therefore cannot all be treated equally. Some form of prioritization is needed before they can be put to use~\cite{DBLP:journals/corr/abs-2607-22219}.

A well-established way to prioritize candidate dependencies is to assign each a strength score and rank candidates accordingly~\cite{DBLP:journals/is/GiannellaR04,DBLP:journals/tcs/KivinenM95,3000767.3000784}. Measures such as $\mu^+$, $\tau$, and $g_3'$ quantify AFD strength from different perspectives, and comparative studies evaluate their behavior and recommend which measures are preferable for discovering or ranking dependencies in noisy data~\cite{DBLP:journals/vldb/ParciakWHNPV25}. Underlying this practice is an implicit assumption: that a dependency ranked higher by strength is also a better candidate for the downstream task the ranking is meant to support.

This paper examines that assumption for error detection, a consequential use of AFDs~\cite{DBLP:series/synthesis/2018Abedjan,DBLP:books/acm/IlyasC19}. We refer to the usefulness of a candidate when deployed as an error detector as its \emph{detection value}. A strength measure characterizes how closely the observed data conform to a dependency, for example through the amount of violation that must be removed or through the concentration of RHS values within LHS equivalence classes~\cite{DBLP:journals/vldb/ParciakWHNPV25}. Detection value concerns something different: how well the candidate avoids false alarms and recovers genuine errors, reflected respectively in \emph{precision} and \emph{recall}. This is a retrieval problem, not a satisfaction problem. Nothing in the definition of these strength measures guarantees that a dependency scoring highly on approximate satisfaction will also have high detection value.

Whether established AFD strength measures actually align with detection value has received limited systematic examination, in part because answering this question requires a notion of value defined by the task itself rather than by the strength measure being evaluated. Rather than proposing another strength measure, we construct a controlled error-detection setting in which the ground-truth errors are known and empirically test the assumption behind strength-based prioritization. Specifically, we ask:
\begin{itemize}
	\item \textbf{RQ1:} Do strength-based rankings reliably track precision in error detection?
	\item \textbf{RQ2:} Do they reliably track recall?
	\item \textbf{RQ3:} If strength and detection value do not consistently align, can the mismatch be explained by a common underlying factor?
\end{itemize}

To answer these questions, we evaluate candidate AFDs directly by their observed error-detection outcomes. We treat detection value as multidimensional rather than defining it through a single scalar score. We report precision and recall together with the corresponding false-positive and false-negative counts, leaving the relative importance of different error types unspecified at the evaluation stage. Strength-based rankings can then be compared separately with rankings induced by these task outcomes, including through top-$k$ overlap. Only when a single candidate-selection objective is required, as in regret analysis, do we introduce an explicit cost over false positives and false negatives. Equal costs serve as a reference setting, and we later test the sensitivity of the conclusions to alternative weightings.

Applying this methodology across three real-world datasets, two corruption mechanisms, and multiple error rates and random seeds, we observe a clear asymmetry under the detection semantics and injection strategies used in the batch experiments. Strength measures tend to track precision positively, although with different degrees of stability. For $\mu^+$, the mean correlation with precision ranges from $0.81$ to $0.88$ across the three datasets. Recall is substantially less stable: both the magnitude and even the direction of the association vary across measures and experimental settings. The mismatch can be pronounced at the candidate level. For example, a dependency with $\mu^+$ above $0.98$ can recover less than $1\%$ of the injected errors. This instability also affects candidate prioritization: top-$k$ sets induced by strength often have little overlap with those induced by actual detection performance, and strength-based selections can incur substantial regret.

An important part of the recall mismatch is associated with structural detection opportunity under the candidate equivalence-class partition: whether an injected error occurs in a region where the candidate has an opportunity to make it observable. This property is not encoded by the strength measures we study, yet it tracks recall more consistently overall across our experiments. At the same time, it does not account for all of the observed variation. We also test corruption mechanism and several structural properties, including RHS cardinality, how often the RHS attribute appears in the LHS of other dependencies, and LHS uniqueness. These factors account for selected effects but do not provide a common explanation across settings. High LHS uniqueness in particular emerges as a severe failure mode rather than a necessary condition for mismatch.

This study makes three main contributions:

\begin{itemize}
\item We develop a task-aligned framework that separates multidimensional error-detection evaluation from explicit cost-based selection.
	
	\item We systematically examine how established AFD strength measures relate to precision, recall, and ranking utility.
	
	\item We analyze the structural sources of recall mismatch, identifying detection opportunity as an important factor and examining alternative explanations for the remaining variation.
\end{itemize}

The remainder of this paper is organized as follows. Sections~\ref{sec:prelim} and~\ref{sec:design} present the preliminaries, evaluation methodology, and experimental design. Sections~\ref{sec:main-results} and~\ref{sec:mismatch} report the main empirical results and investigate the sources of the observed mismatch. Sections~\ref{sec:discussion}--\ref{sec:related} discuss implications, robustness, scope, and prior work, and Section~\ref{sec:conclusion} concludes.

\section{Preliminaries and Evaluation Methodology}
\label{sec:prelim}

This section establishes the preliminaries and evaluation methodology used throughout the paper. We first define AFD candidates, the tuple partitions on which their strength measures are computed, and the three strength measures considered in this study. We then formalize task-aligned detection value and introduce a decision-oriented cost for analyses that require a single candidate-selection objective. Table~\ref{tab:notation} summarizes the main notation.

\begin{table*}[t]
	\centering
	\caption{Notation used throughout this paper.}
	\label{tab:notation}
	\begin{tabular}{ll}
		\toprule
		Symbol & Meaning \\
		\midrule
		$R$, $r$ & relation schema; instance of $R$ \\
		$u$ & corruption-and-detection run \\
		$X \to A$ & candidate FD, LHS $X$, RHS $A$ \\
		$L$ & maximum LHS size \\
		$\phi$ & shorthand for candidate $X \to A$ \\
		$\Pi_X$ & equivalence-class partition induced by $X$ over tuples non-null on $X$ \\
		$r_\phi$ & subrelation of tuples non-null on $X \cup \{A\}$ \\
		$N_\phi$ & number of tuples in $r_\phi$ \\
		$K_\phi$ & number of nonempty $X$-classes after restriction to $r_\phi$ \\
		$n_{c,a}$ & tuples in class $c$ with $A$-value $a$ \\
		$n_c$, $n_a$ & class size; global frequency of value $a$ \\
		$\mathrm{pdep}(X,A)$, $\mathrm{pdep}(A)$ & probabilistic dependency of $X$ on $A$; of $A$ alone \\
		$\tau$, $\mu$, $\mu^+$, $g_3$, $g_3'$ & quantities used to define AFD strength (Sec.~\ref{sec:prelim-afd}) \\
		$E$ & cells altered by injection in a run \\
		$E_{rel}(\phi)$ & cells in $E$ relevant to $\phi$ \\
		$E_{vis}(\phi)$ & structurally visible subset of $E_{rel}(\phi)$ \\
		$F_\phi$ & cells flagged by a detector for $\phi$ \\
		$TP_\phi$, $FP_\phi$, $FN_\phi$ & true positives, false positives, false negatives \\
		$\mathrm{precision}_\phi$, $\mathrm{recall}_\phi$, $\mathrm{recall}_{vis,\phi}$ & precision; default and visible-only recall \\
		$V_{use}(\phi)$ & four-component detection-value vector for $\phi$ \\
		$L_{detect}(\phi)$ & weighted detection cost \\
		$c_{FN}$, $c_{FP}$ & weights on FN and FP in $L_{detect}$ \\
		$\lambda$ & cost ratio $c_{FN}/c_{FP}$ \\
		$\phi^*$ & candidate minimizing $L_{detect}$ among competing candidates \\
		$\hat\phi$ & candidate ranked first by a given strength measure among competing candidates \\
		$\mathrm{regret}(\hat\phi)$, $\mathrm{regret}_{norm}(\hat\phi)$ & absolute and normalized regret of the top-ranked candidate \\
		$\rho_{ns}(\phi)$ & fraction of candidate-valid rows in non-singleton classes \\
		$\rho_{vis}(\phi)$ & visibility ratio, $|E_{vis}(\phi)|/|E_{rel}(\phi)|$ \\
		$q_{ns}^{(u)}(\phi)$ & fraction of relevant errors in run $u$ that fall on non-singleton rows \\
		\bottomrule
	\end{tabular}
\end{table*}

\subsection{AFDs and Strength Measures}
\label{sec:prelim-afd}

Let $R$ be a relation schema with attribute set $\mathrm{attr}(R)$, and let $r$ be an instance of $R$. For an attribute set $X \subseteq \mathrm{attr}(R)$ and a single attribute $A \in \mathrm{attr}(R) \setminus X$, a functional dependency $X \to A$ asserts that every pair of tuples in $r$ agreeing on $X$ also agrees on $A$. We refer to $X$ as the left-hand side (LHS) and $A$ as the right-hand side (RHS). An \emph{approximate} functional dependency (AFD) relaxes exact satisfaction: $X \to A$ may be violated by some tuples, and a strength measure quantifies how closely the observed data satisfy the dependency. We restrict the candidate space to $1 \le |X| \le L$ for a fixed bound $L$. This bound limits the search space rather than the definition of an AFD itself. Section~\ref{sec:design-detection} describes how $L$ is chosen.

For a fixed LHS $X$, tuples with the same non-null values on all attributes in $X$ belong to the same \emph{equivalence class}. Tuples containing a null on any attribute in $X$ are excluded from these classes. The collection of the resulting equivalence classes forms the partition $\Pi_X$. Because $\Pi_X$ depends only on $X$, all candidates sharing the same LHS use the same underlying partition, regardless of their RHS attribute.

Now consider a specific candidate $\phi : X \to A$. We define its valid subrelation as
\[
r_\phi
=
\{\,t \in r
\mid
t[B] \neq \mathrm{NULL}
\text{ for every } B \in X \cup \{A\}
\,\}.
\]
Thus, $r_\phi$ contains exactly the tuples on which both the LHS and RHS of $\phi$ are defined. Candidate-specific equivalence classes are obtained by restricting each class in $\Pi_X$ to tuples in $r_\phi$ and discarding empty intersections. For convenience, we continue to denote each resulting nonempty class by $c$. We write
$
N_\phi = |r_\phi|
$
for the number of candidate-valid tuples and $K_\phi$ for the number of resulting nonempty equivalence classes. Within each class $c$, the distribution of $A$-values determines whether and to what extent $X \to A$ is violated: a class containing only one $A$-value satisfies the dependency internally, whereas a class containing multiple $A$-values contributes a violation. All class-based quantities and sums below are computed over these $K_\phi$ candidate-specific classes.

The ratio
$
\mathrm{lhsUniqueness}(\phi) = \frac{K_\phi}{N_\phi}
$
summarizes how finely $X$ partitions the candidate-valid tuples in $r_\phi$. It approaches $0$ when most tuples are concentrated in a small number of large classes and equals $1$ when every class is a singleton, in which case $X$ acts as a key on $r_\phi$. The quantity is undefined when $N_\phi = 0$.

For each candidate-specific equivalence class $c$, let $n_{c,a}$ denote the number of tuples in $c$ whose $A$-value is $a$. We write
$
n_c = \sum_a n_{c,a}
$
for the size of class $c$, and
$
n_a = \sum_c n_{c,a}
$
for the number of tuples in $r_\phi$ whose $A$-value is $a$. These counts form the basis of the probabilistic dependency quantities used below. The \emph{probabilistic dependency} of $X$ on $A$~\cite{3000767.3000784} is defined as
\[
\mathrm{pdep}(X, A)
\;=\;
\frac{1}{N_\phi}
\sum_c \sum_a
\frac{n_{c,a}^{2}}{n_c}.
\]
Within each class, the term
$
\sum_a \frac{n_{c,a}^{2}}{n_c}
$
increases as the distribution of $A$ becomes more concentrated. Consequently, $\mathrm{pdep}(X,A)$ reaches $1$ when every candidate-specific class is homogeneous in $A$.

The corresponding quantity for $A$ alone ignores the partition induced by $X$ and is defined as
\[
\mathrm{pdep}(A)
\;=\;
\sum_a
\left(
\frac{n_a}{N_\phi}
\right)^2.
\]
It measures the concentration of $A$ over the same candidate-valid tuples in $r_\phi$. Because $\mathrm{pdep}(X,A)$ and $\mathrm{pdep}(A)$ are computed on the same set of tuples, the latter provides a directly comparable baseline for the former.

The three strength measures evaluated in this paper are $\tau$, $\mu^+$, and $g_3'$. The measures $\tau$ and $\mu$ were introduced by Piatetsky-Shapiro and Matheus~\cite{3000767.3000784} in the probabilistic-dependency framework. We use the non-negative variant $\mu^+$ of $\mu$, following Parciak et al.~\cite{DBLP:journals/vldb/ParciakWHNPV25}. The measure $g_3'$~\cite{DBLP:journals/is/GiannellaR04} is a normalized variant of the violation-based measure $g_3$, introduced by Kivinen and Mannila~\cite{DBLP:journals/tcs/KivinenM95}. The definitions of these measures are given below.

The measure $\tau$ compares the concentration of $A$ within the classes induced by $X$ with the concentration of $A$ observed globally on the same candidate-valid tuples:
\[
\tau
\;=\;
\frac{\mathrm{pdep}(X,A)-\mathrm{pdep}(A)}
{1-\mathrm{pdep}(A)}.
\]
Thus, $\mathrm{pdep}(A)$ serves as the baseline against which the additional concentration associated with $X$ is evaluated.

The measure $\mu$ uses a chance-adjusted baseline instead. It compares the observed $\mathrm{pdep}(X,A)$ with the expected value of this quantity under a no-dependence model that preserves the number of candidate-valid tuples and the number of equivalence classes:
\[
\mathbb{E}[\mathrm{pdep}(X,A)]
\;=\;
\mathrm{pdep}(A)
+
\frac{K_\phi-1}{N_\phi-1}
\bigl(1-\mathrm{pdep}(A)\bigr).
\]
The adjustment accounts for the fact that a finer partition can exhibit stronger apparent concentration even in the absence of a genuine dependence between $X$ and $A$. Using this expectation as the baseline,
\[
\mu
\;=\;
\frac{\mathrm{pdep}(X,A)-\mathbb{E}[\mathrm{pdep}(X,A)]}
{1-\mathbb{E}[\mathrm{pdep}(X,A)]}.
\]
Because $\mu$ may become negative when the observed partition performs worse than this baseline, we use its non-negative variant
\[
\mu^+
\;=\;
\max(\mu,0).
\]

The measure $g_3'$ is defined through $g_3$, which takes a different view of approximate satisfaction. For each equivalence class, $g_3$ retains the largest group of tuples sharing the same $A$-value:
\[
g_3
\;=\;
\frac{1}{N_\phi}
\sum_c
\max_a n_{c,a}.
\]
Accordingly, $1-g_3$ is the minimum fraction of candidate-valid tuples that must be removed for $X \to A$ to hold exactly. The normalized measure $g_3'$ corrects $g_3$ for the one-tuple-per-class agreement that is unavoidable in any partition:
\[
g_3'
\;=\;
\frac{N_\phi g_3-K_\phi}
{N_\phi-K_\phi}.
\]

Following the classification of Parciak et al.~\cite{DBLP:journals/vldb/ParciakWHNPV25}, we refer to $\tau$ and $\mu^+$ as \textsc{Logical} measures and to $g_3'$ as a \textsc{Violation} measure.

Each candidate falls into one of three mutually exclusive cases.

\begin{itemize}
	\item \emph{Undefined}. If $N_\phi = 0$, then $r_\phi$ contains no candidate-valid tuple and none of the three strength measures is defined. We therefore leave $\tau$, $\mu^+$, and $g_3'$ undefined rather than assigning them an arbitrary default value.
	
	\item \emph{Exact satisfaction}. Suppose $N_\phi > 0$ and $X \to A$ holds exactly on $r_\phi$. Every candidate-specific equivalence class is homogeneous in $A$, so $\mathrm{pdep}(X,A)=1$. We assign
	$
	\tau=\mu^+=g_3'=1.
	$
	For ordinary exact cases, this value is obtained directly from the formulas above. The convention also covers two degenerate exact cases. If $A$ is globally constant on $r_\phi$, then $\mathrm{pdep}(A)=1$, causing the formulas for $\tau$ and $\mu$ to reduce to $0/0$. If $X$ is a key on $r_\phi$, then $K_\phi=N_\phi$, causing the formulas for $\mu$ and $g_3'$ to reduce to $0/0$. When both conditions hold, all three formulas are degenerate, and we set $\tau=\mu^+=g_3'=1$ by the convention above.
	
	\item \emph{General case}. In every non-exact case with $N_\phi>0$, at least one equivalence class contains multiple $A$-values. Consequently, $\mathrm{pdep}(A)<1$ and $K_\phi<N_\phi$, so the formulas for $\tau$, $\mu$, and $g_3'$ are well defined and are applied directly.
\end{itemize}

These conventions define the three strength measures for every candidate with $N_\phi>0$, while candidates with $N_\phi=0$ remain undefined.

\subsection{Task-Aligned Detection Value}
\label{sec:prelim-vuse}

While Section~\ref{sec:prelim-afd} characterizes how strongly an AFD $\phi$ holds in the observed data, using an AFD for error detection requires evaluating its performance as a detection rule. We refer to this task-specific performance as the \emph{detection value} of the candidate and summarize it later by $V_{use}(\phi)$.

Detection is evaluated against injected errors whose locations are known. In each corruption run, selected cells are altered to create a corrupted instance, and we let $E$ denote the set of all altered cells. Ground truth is therefore defined at the cell level: an error corresponds to a specific $(\text{row},\text{attribute})$ pair rather than to an entire tuple.

For a candidate $\phi:X\to A$, only injected errors on its RHS attribute $A$ and within its candidate-valid tuples are relevant to its detection performance. We denote this subset by
\[
E_{rel}(\phi) \subseteq E.
\]
Errors on other attributes or outside the tuples considered by $\phi$ are not counted against the candidate. Relevance, however, does not imply detectability. Some errors in $E_{rel}(\phi)$ may be structurally impossible for $\phi$ to expose.

We therefore define $E_{vis}(\phi) \subseteq E_{rel}(\phi)$ as the subset of \emph{structurally visible} errors. A relevant error is structurally visible if its row belongs to a conflicting candidate-specific equivalence class after corruption. A class is conflicting if it contains more than one candidate-valid row and more than one distinct RHS value. This definition depends only on the post-corruption class structure. It does not require the class to satisfy the dependency before corruption, and it does not require a unique majority.

Given a candidate $\phi$, let $F_\phi$ denote the set of cells flagged as suspicious when the detector is applied to the corrupted instance. The detection semantics that determine $F_\phi$ are specified in Section~\ref{sec:design-detection}. Relative to the candidate-relevant injected errors $E_{rel}(\phi)$, we define the \emph{true-positive} set as
\[
TP_\phi
=
F_\phi \cap E_{rel}(\phi),
\]
the \emph{false-positive} set as
\[
FP_\phi
=
F_\phi \setminus E_{rel}(\phi),
\]
and the \emph{false-negative} set as
\[
FN_\phi
=
E_{rel}(\phi) \setminus F_\phi.
\]
Precision is then
\[
\mathrm{precision}_\phi
=
\frac{|TP_\phi|}
{|TP_\phi|+|FP_\phi|}.
\]

Our default recall is evaluated against all candidate-relevant errors:
\[
\mathrm{recall}_\phi
=
\frac{|TP_\phi|}
{|E_{rel}(\phi)|}.
\]
Structurally invisible errors remain in the denominator. This choice treats the inability of a candidate to expose such errors as part of its detection performance rather than removing them from evaluation.

For diagnostic purposes, we also report recall restricted to structurally visible errors:
\[
\mathrm{recall}_{vis,\phi}
=
\frac{|F_\phi \cap E_{vis}(\phi)|}
{|E_{vis}(\phi)|}.
\]
This quantity measures detector performance conditional on an error being structurally visible and is not used as the default recall.

Separately, we define the \emph{visibility ratio}
\[
\rho_{vis}(\phi)
=
\frac{|E_{vis}(\phi)|}
{|E_{rel}(\phi)|}.
\]
It measures the fraction of candidate-relevant errors whose rows belong to conflicting classes after corruption. A low value indicates that many relevant errors do not become structurally visible, independently of how successfully the detector handles the errors that are visible. We return to this distinction in Section~\ref{sec:mismatch}.

Collecting these quantities, we represent the detection value of a candidate $\phi$ by the four-component vector
\[
V_{use}(\phi)
\;=\;
\bigl(
\mathrm{precision}_\phi,
\mathrm{recall}_\phi,
|FP_\phi|,
|FN_\phi|
\bigr).
\]
The first two components summarize relative detection performance, while the latter two retain the corresponding absolute numbers of false alarms and missed errors. We do not include $|TP_\phi|$ separately because its contribution is already reflected in precision and recall. The diagnostic quantities $\mathrm{recall}_{vis,\phi}$ and $\rho_{vis}(\phi)$ are kept outside $V_{use}(\phi)$ because they are used to analyze why recall is high or low rather than to summarize the overall detection performance of the candidate.

We treat $V_{use}(\phi)$ as a multidimensional evaluation rather than reducing it to a single scalar at this stage. Different applications may assign different importance to false alarms and missed errors, so the evaluation itself does not impose such a preference. When a subsequent analysis requires a single candidate-selection objective, Section~\ref{sec:prelim-ldetect} introduces an explicit cost based on $|FP_\phi|$ and $|FN_\phi|$.

\subsection{Decision-Oriented Cost}
\label{sec:prelim-ldetect}

Some analyses require a single candidate-selection objective rather than a multidimensional description of detection performance. In particular, regret analysis requires identifying which candidate is preferable under an explicit notion of detection cost. For this purpose, we define a scalar loss from the false-positive and false-negative counts already included in $V_{use}(\phi)$.

Given a candidate $\phi$, we define
\[
L_{detect}(\phi)
\;=\;
c_{FN}\,|FN_\phi|
+
c_{FP}\,|FP_\phi|,
\]
where $c_{FN}$ is the cost of a missed error and $c_{FP}$ is the cost of a false alarm. We construct the loss directly from $|FN_\phi|$ and $|FP_\phi|$ because these counts correspond to the two decision costs of interest. Precision and recall remain normalized evaluation summaries derived from the same detection outcomes and are therefore not added separately to the loss.

The false-negative count is defined with respect to $E_{rel}(\phi)$, consistent with the default recall in Section~\ref{sec:prelim-vuse}. Thus, a structurally invisible relevant error that remains undetected still contributes to $L_{detect}(\phi)$: it is a genuine error that the candidate fails to surface, regardless of the reason for that failure. Lower values of $L_{detect}$ indicate lower detection cost under the specified weighting.

The weights $c_{FN}$ and $c_{FP}$ determine the trade-off between missed errors and false alarms. This trade-off depends on the deployment context, so we do not treat any particular weighting as intrinsically correct. We use
\[
c_{FN}=c_{FP}=1
\]
as a reference setting in the main analysis and examine the sensitivity of the results to alternative cost ratios in Section~\ref{sec:threats}.

When a strength measure is used to select a single candidate, ranking disagreement matters not only in whether the selected candidate differs from the cost-minimizing one, but also in how much additional detection cost that choice incurs. Consider candidates that share the same RHS attribute and therefore provide alternative detection rules for errors on that attribute. Let $\phi^*$ denote a candidate minimizing $L_{detect}$ under the specified cost weighting, and let $\hat{\phi}$ denote the candidate ranked first by a given strength measure. We define the \emph{regret} of this selection as
\[
\mathrm{regret}(\hat{\phi})
\;=\;
L_{detect}(\hat{\phi})
-
L_{detect}(\phi^*).
\]
A regret of zero means that the top-ranked candidate under the strength measure is also cost-minimizing. Larger regret indicates a greater additional detection cost from following the strength-based selection.

Because the scale of $L_{detect}$ can vary substantially across RHS attributes, the same absolute regret may have very different significance across candidate sets. We therefore also report \emph{normalized regret},
\[
\mathrm{regret}_{norm}(\hat{\phi})
\;=\;
\frac{\mathrm{regret}(\hat{\phi})}
{\max\bigl(1,L_{detect}(\phi^*)\bigr)}.
\]
This normalization expresses the additional cost relative to the cost of the best available candidate, while flooring the denominator at $1$ to keep the ratio well defined and stable when the minimum cost is zero or very small.

\section{Experimental Design}
\label{sec:design}

\subsection{Datasets and Measures}
\label{sec:datasets}

Our experiments cover the \textsc{Violation} and \textsc{Logical} families identified by Parciak et al.~\cite{DBLP:journals/vldb/ParciakWHNPV25}, represented by $g_3'$ and by $\tau$ and $\mu^+$, respectively. We do not include the \textsc{Shannon} family because its computational cost is incompatible with the scale of our batch experiments. Representative measures in this family include $RFI^+$ and its normalized variant $RFI'^+$~\cite{DBLP:conf/kdd/MandrosBV17}. Parciak et al. report that, under a 24-hour runtime limit on a single CPU core, these measures compute scores for only 250 of 1,634 candidate FDs on their benchmark. In comparison, our complete experimental pipeline, including candidate scoring and all corruption-and-detection runs, completes in approximately 16.5 hours using $\mu^+$, $\tau$, and $g_3'$. Extrapolating the reported $RFI^+$/$RFI'^+$ scoring cost to the scale of our study would require several years of computation. We therefore restrict the main comparison to the two measure families that can be evaluated at the required scale.

\begin{table*}[t]
	\centering
	\caption{Selected attributes that recur in later structural analyses. Cardinality is the number of distinct non-null values out of the total row count of the relation.}
	\label{tab:attribute-roles}
	\begin{tabular}{llll}
		\toprule
		Attribute & Dataset & Role & Cardinality \\
		\midrule
		fnlwgt & Adult & high-cardinality, fragments equivalence classes & $21{,}648/32{,}561$ \\
		Claim Number & Claims & near-key, uniquely identifies almost every row & $97{,}230/97{,}231$ \\
		Condition & Hospital & low-cardinality, yields a coarse LHS partition & $5/114{,}919$ \\
		\bottomrule
	\end{tabular}
\end{table*}

We conduct our experiments on three real-world relations: Adult ($32{,}561$ rows, 15 attributes), Hospital ($114{,}919$ rows, 15 attributes), and Claims ($97{,}231$ rows, 13 attributes).\footnote{All three relations are available from the benchmark of Parciak et al.~\cite{parciak_2023_8098909}, DOI: 10.5281/zenodo.8098909.} Table~\ref{tab:attribute-roles} lists selected attributes whose structural properties recur in later analyses.

The main experiments consider candidates with $1 \le |X| \le 3$. This bound keeps candidate enumeration and repeated detection evaluation tractable at batch scale. Section~\ref{sec:threats} later examines the sensitivity of the main conclusions to extending the search to $|X| \le 4$. For each relation, we define its \emph{target pool} as the set of minimal exact FDs within this candidate space. These dependencies serve as the targets for FD-aligned error injection.

The three relations provide target pools with different structural characteristics. Adult has the smallest target pool, containing only 2 minimal exact FDs, and therefore provides a relatively simple setting in which candidate-level behavior can be inspected closely. Hospital has a much richer target pool of 48 minimal FDs spanning a wide range of RHS cardinalities, allowing the analysis to cover substantially more heterogeneous dependency structures. Claims contains 8 minimal exact FDs and provides a third relation with a different structural profile from the other two.

Adult, Claims, and Hospital are drawn from the real-world-data benchmark used by Parciak et al. to evaluate AFD measures~\cite{DBLP:journals/vldb/ParciakWHNPV25}. They correspond to R1, R2, and R4 in that benchmark, with matching row and attribute counts. Using relations from the same benchmark allows us to evaluate the downstream detection behavior of the strength measures in a setting already used for their comparative study.

We additionally cross-check our exact-FD extraction against the manually annotated design dependencies released with the benchmark~\cite{parciak_2023_8098909}. Because these annotations are available for linear dependencies, we restrict this check to $|X|=1$. Every dependency classified as an exact PFD in the benchmark is recovered by our independent extraction as an exact minimal FD: 2 of 2 on Adult, 2 of 2 on Claims, and 22 of 22 on Hospital. This agreement provides an independent check on the construction of the exact-FD target pools used in our experiments.

\subsection{Corruption Models}
\label{sec:corruption}

We use two corruption mechanisms. \emph{FD-aligned injection} modifies the RHS of a target dependency according to its equivalence-class structure, whereas \emph{general corruption} perturbs a single attribute without reference to any candidate dependency. Using both mechanisms allows us to assess whether observed patterns are specific to the dependency-aware construction of injected errors or also arise under corruption generated independently of the candidate being evaluated. In particular, observing the same pattern under general corruption reduces the possibility that it is induced by the FD-aligned injection procedure itself.

We refer to the full set of automated runs over all FD targets for FD-aligned corruption and all distinct RHS attributes represented in each target pool for general corruption, together with multiple error rates and random seeds, as the \emph{batch experiments}. Their results are analyzed starting in Section~\ref{sec:main-results}.

\emph{FD-aligned injection} modifies only the RHS of a target candidate $\phi:X\to A$ and leaves all attributes in $X$ unchanged. Altering an LHS attribute would change the equivalence-class partition itself, making it difficult to separate the effect of introducing an RHS violation from the effect of changing the grouping structure on which the dependency is evaluated. Keeping $X$ fixed avoids this confounding.

Injection is restricted to cells in non-singleton equivalence classes. Changing the RHS value of a tuple in a singleton class cannot create a within-class conflict, because there is no other tuple in the class with which it can disagree. Within each eligible class, at most half of its tuples are corrupted. Because the target dependency holds exactly before injection, this cap prevents corrupted tuples from constituting a strict majority within the class. Without such a cap, sufficiently extensive corruption could redefine the dominant value of the class rather than create a minority of anomalous values. We generate replacement values under three strategies.
\begin{itemize}
	\item \texttt{WITHIN\_DOMAIN} draws a replacement uniformly at random from the observed domain of the attribute, excluding the original value.
	\item \texttt{CROSS\_CLASS} copies a value observed in a different equivalence class on the same attribute. The selected value is required to differ from the original value. If repeated cross-class sampling fails to find such a value, the injector falls back to a different value from the observed domain. The replacement therefore remains an observed value while ensuring that the target cell is actually changed.
	\item \texttt{FRESH\_VALUE} introduces a value that does not previously occur in the observed domain of the attribute.
\end{itemize}

\emph{General corruption} targets one attribute directly, without using any candidate dependency or equivalence-class structure. Every non-null cell on the selected attribute is eligible, and no per-class corruption cap is imposed. Each run corrupts only one attribute. Perturbing multiple attributes in the same run could modify both the LHS and RHS of a candidate, making it difficult to separate changes in the equivalence-class structure from changes in the RHS values of the candidate. Restricting each run to one attribute avoids this confounding. In the batch experiments, replacement values are generated by \texttt{RANDOM\_COPY}, which copies a value from another randomly selected non-null cell on the same attribute while requiring the replacement to differ from the original value. Because sampling is performed over cells rather than distinct values, more frequent values are more likely to be selected. This parallels \texttt{CROSS\_CLASS} while remaining independent of any equivalence-class structure.

For FD-aligned corruption, the batch experiments exhaustively cover all targets in the three target pools: 2 minimal exact FDs for Adult, 8 for Claims, and 48 for Hospital. The larger Hospital pool spans every attribute that appears as the RHS of a minimal FD and covers a wide range of RHS cardinalities, from \textit{EmergencyService} with cardinality 2 to \textit{ProviderNumber} with cardinality 4,598. This coverage allows us to examine the relationship between RHS cardinality and strength-detection behavior without selecting targets on that basis. The analysis is presented in Section~\ref{sec:mismatch}.

Across all three datasets, every FD-aligned target and every general-corruption RHS attribute uses the same $5 \times 3$ grid of random seeds $\{1,7,42,100,123\}$ and error rates $\{0.01,0.05,0.10\}$. Using the same five seeds across all settings provides consistent coverage of injection randomness, while the three error rates represent increasing levels of corruption from $1\%$ to $10\%$.

\texttt{CROSS\_CLASS} is used for FD-aligned corruption at batch scale, while \texttt{RANDOM\_COPY} is used for general corruption. Both strategies use replacement values already observed on the same attribute. \texttt{CROSS\_CLASS} additionally draws the replacement from a different equivalence class, whereas \texttt{RANDOM\_COPY} ignores equivalence-class structure. Thus, both strategies use existing values, but only \texttt{CROSS\_CLASS} uses dependency structure to guide replacement.

The alternative strategies, including \texttt{WITHIN\_DOMAIN} and \texttt{FRESH\_VALUE} are evaluated in small-scale validation runs and in a targeted sensitivity analysis on a subset of targets. These experiments assess whether the main detection-level conclusions depend on the replacement strategy, but do not provide full-scale coverage of every target pool. Section~\ref{sec:threats} reports the corresponding sensitivity results and discusses this scope limitation.

\subsection{Ground Truth and Detection Semantics}
\label{sec:design-detection}

The evaluation quantities $E$, $E_{rel}(\phi)$, $E_{vis}(\phi)$, precision, and recall are defined in Section~\ref{sec:prelim-vuse}. Here we specify how the flagged set $F_\phi$ is produced. We consider two detection semantics, \textsc{Non-Majority} and \textsc{Whole-Group}. Under \textsc{Non-Majority}, only tuples whose $A$-value occurs less frequently than the unique majority value of a conflicting equivalence class are flagged. If no value has a unique highest frequency, no tuple in the class is flagged. Under \textsc{Whole-Group}, every tuple in a conflicting class is flagged.

The \textsc{Non-Majority} semantics follows a common principle in constraint-based cleaning: a violation identifies a conflicting group, while within-group evidence such as value frequency or minimum-change reasoning is used to localize the values or cells more likely to be erroneous~\cite{DBLP:conf/sigmod/BohannonFFR05,DBLP:conf/sigmod/DallachiesaEEEIOT13,DBLP:journals/tods/FanGJK08}. We use \textsc{Non-Majority} as the primary semantics because it evaluates whether a candidate can localize suspicious RHS cells within a conflicting class, rather than treating every tuple in that class as equally suspicious.

\textsc{Non-Majority} is used throughout the batch experiments. \textsc{Whole-Group} is evaluated in a small-scale paired comparison on the same corrupted instances and candidate sets, reported in Section~\ref{sec:threats-semantics}. Under \textsc{Whole-Group}, every tuple in a conflicting class is flagged regardless of majority status, whereas \textsc{Non-Majority} flags only values below a unique majority. The paired comparison shows that changing the detection semantics changes the observed relationships between AFD strength and detection performance. The batch experiments therefore remain focused on \textsc{Non-Majority}, and we treat detection semantics as part of the scope of the results.

Each corruption-and-detection run is linked, through a shared run identifier, to its configuration, candidate strength scores, and candidate-level detection outcomes. This allows the full result set to be reconstructed and re-aggregated without rerunning the experiments. In total, the batch experiments comprise 1,155 runs: 60 on Adult, 930 on Hospital, and 165 on Claims, yielding 513,480 candidate-level detection evaluations.

\subsection{Run-Level Correlation}
\label{sec:run-level-correlation}

A main-experiment run $u$ is identified by its dataset, target, injection mechanism, error rate, and random seed. The replacement strategy is fixed by the corruption mechanism as described in Section~\ref{sec:corruption}. Within each run, let $C_u$ denote the set of all candidates in the main candidate space that share the same RHS attribute as the target.

For each strength measure and each detection metric $Y\in\{\mathrm{precision},\mathrm{recall}\}$, where $Y(\phi)$ denotes the value of that metric for candidate $\phi$, we compute one run-level Spearman rank correlation
\[
\rho_u^{(Y)}
=
\rho_{\mathrm{S}}
\left(
\{S(\phi)\}_{\phi\in C_u},
\{Y(\phi)\}_{\phi\in C_u}
\right),
\]
where $\rho_{\mathrm{S}}$ denotes the Spearman rank correlation coefficient and $S(\phi)$ is the strength score of candidate $\phi$. Ties are handled using the standard average-rank convention~\cite{hollander2014nonparametric}, since identical precision or recall values occur frequently in our data.

If all candidates in a run have the same strength score or the same value for the detection metric, the corresponding Spearman correlation is undefined. In either case, one of the two ranked variables has zero variance. Such runs are excluded from dataset-level aggregation rather than assigned a correlation of zero. In practice, no run in the main experiments triggers this exclusion. For dataset $D$, let $\mathcal{U}_D^{(Y)}$ denote the set of runs for which $\rho_u^{(Y)}$ is defined. We report the unweighted arithmetic mean
\[
\bar{\rho}_D^{(Y)}
=
\frac{1}{|\mathcal{U}_D^{(Y)}|}
\sum_{u\in\mathcal{U}_D^{(Y)}} \rho_u^{(Y)}.
\]
The pooled correlation reported in Section~\ref{sec:main-results} is computed analogously over all runs with a defined correlation, regardless of dataset.

This run definition applies to the main batch experiments. The replacement-strategy sensitivity analysis in Section~\ref{sec:threats} varies the replacement strategy independently and is analyzed separately from the main batch experiments.

\section{Does Strength Track Detection Value?}
\label{sec:main-results}

This section asks whether AFD strength tracks detection value, the central question of this paper. We represent detection value by the four-component vector
\[
V_{use}=(\mathrm{precision},\mathrm{recall},|FP|,|FN|),
\]
rather than by a single scalar. For strength to align with detection value in a way that supports candidate selection, its relationship with these task-specific outcomes should be sufficiently stable.

We examine this question from three complementary perspectives. Sections~\ref{sec:precision} and~\ref{sec:recall} study precision and recall directly. Precision asks how often the cells flagged by a candidate are genuine errors, whereas recall asks how many of the relevant errors are detected.

Section~\ref{sec:ranking-utility} examines what happens when strength is used to rank candidates for detection. A positive association with a detection metric does not guarantee good candidate ranking. Candidates ranked highest by strength may still be suboptimal for the downstream task. We evaluate this ranking utility in three ways:
\begin{itemize}
	\item We compare the rankings induced by strength with those induced by precision and recall.
	\item We use the cost-oriented objective $L_{detect}$ to combine $|FP|$ and $|FN|$ under an explicit cost weighting, and measure the regret incurred when the top-ranked candidate by strength is selected instead of the candidate minimizing detection cost.
	\item We examine cases in which different strength measures select the same top-ranked candidate, and ask whether such agreement corresponds to high detection value or instead arises from structural properties of the candidate space.
\end{itemize}

\subsection{Precision}
\label{sec:precision}

We first examine whether higher AFD strength is associated with higher detection precision. Table~\ref{tab:precision-correlation} reports the mean run-level Spearman correlations defined in Section~\ref{sec:run-level-correlation}. The association is consistently positive across the three datasets, although its magnitude varies across measures. For $\mu^+$, which Parciak et al. recommend for practical AFD discovery, the mean correlation ranges from $0.8115$ to $0.8827$. Thus, candidates ranked higher by $\mu^+$ tend to have higher precision, meaning that a larger fraction of their flagged cells are genuine errors rather than false positives. Because Hospital contributes 930 of the 1,155 runs, compared with 60 for Adult and 165 for Claims, the pooled mean is weighted more heavily toward Hospital and should not be interpreted as an equal-weight average over the three datasets.

\begin{table}[t]
	\centering
	\caption{Mean run-level Spearman correlation between strength and precision, by measure and dataset. The final column pools all 1155 runs across datasets and corruption mechanisms into a single average, weighting each run equally regardless of dataset.}
	\label{tab:precision-correlation}
	\begin{tabular}{lcccc}
		\toprule
		Measure & Adult & Hospital & Claims & Pooled \\
		\midrule
		$\mu^+$   & 0.8115 & 0.8827 & 0.8636 & 0.8763 \\
		$g_3'$    & 0.7067 & 0.8525 & 0.8732 & 0.8479 \\
		$\tau$    & 0.6971 & 0.6527 & 0.5805 & 0.6447 \\
		\bottomrule
	\end{tabular}
\end{table}

The positive association with precision is not unique to $\mu^+$. $g_3'$ also shows strong positive correlations, with a pooled mean of $0.8479$ and high positive means on all three datasets. In contrast, $\tau$ is consistently weaker: its mean correlation remains positive on every dataset but falls to $0.5805$ on Claims, with a pooled mean of $0.6447$. On Adult, the mean correlation changes by roughly $0.1$ between FD-aligned and general corruption for both $\tau$ and $g_3'$. No comparable difference between the two corruption mechanisms appears on Hospital or Claims. We return to the possible role of corruption mechanism in Section~\ref{sec:mismatch}.

The strong mean correlations do not imply uniformly high correlations across individual runs. Across all three measures, the run-level correlations range from $-0.0111$ to $0.9886$. The only negative strength-precision correlation in the batch experiments is $-0.0111$, observed for $\tau$ in one Hospital run. The lower values are concentrated in $\tau$ on Hospital, where the correlation decreases as the error rate increases. This pattern differs from the corruption-mechanism difference noted above, which compares FD-aligned and general corruption rather than error rates. We do not further investigate the source of this measure-specific sensitivity here. The dataset means should therefore be interpreted as overall tendencies rather than as evidence that strength tracks precision equally well in every run.

Precision reflects how often a flagged cell is actually erroneous. The consistently positive association observed here, strongest for $\mu^+$ and $g_3'$ and weaker for $\tau$, indicates that AFD strength is meaningfully associated with false-alarm behavior. Recall provides a complementary test. If strength were to align consistently with detection value, we would expect a similarly stable association with the ability to recover relevant errors.

\subsection{Recall}
\label{sec:recall}

Recall here refers to $\mathrm{recall}_\phi$ as defined in Section~\ref{sec:prelim-vuse}. Section~\ref{sec:precision} shows that strength is positively associated with precision across all three datasets. The corresponding relationship with recall is much less stable. We examine the overall pattern across measures and datasets, variation across RHS attributes within Hospital, and extreme individual candidates that aggregate summaries can conceal.

\subsubsection{Overall Recall Patterns}
\label{sec:recall-aggregate}

Table~\ref{tab:recall-correlation} reports the mean run-level Spearman correlations between strength and recall. The three measures show markedly different behavior. For $\mu^+$, the mean correlation is close to zero on Adult ($0.0129$), weakly positive on Hospital ($0.1239$), and substantially negative on Claims ($-0.4869$). For $g_3'$, the corresponding means are positive on Adult ($0.5688$) and Hospital ($0.2669$), but negative on Claims ($-0.3756$). In contrast, $\tau$ is negatively associated with recall on all three datasets, with mean correlations of $-0.2285$, $-0.5056$, and $-0.7312$ on Adult, Hospital, and Claims, respectively.

\begin{table}[t]
	\centering
	\caption{Mean run-level Spearman correlation between strength and recall, by measure and dataset. The final column pools all runs across datasets and corruption mechanisms, weighting each run equally regardless of dataset.}
	\label{tab:recall-correlation}
	\begin{tabular}{lcccc}
		\toprule
		Measure & Adult & Hospital & Claims & Pooled \\
		\midrule
		$\mu^+$   & 0.0129  & 0.1239  & -0.4869 & 0.0309 \\
		$g_3'$    & 0.5688  & 0.2669  & -0.3756 & 0.1908 \\
		$\tau$    & -0.2285 & -0.5056 & -0.7312 & -0.5234 \\
		\bottomrule
	\end{tabular}
\end{table}

Across individual runs, the variation is even larger. The smallest strength-recall correlation is $-0.8267$, for $\tau$ on a Claims run, whereas the largest is $0.8867$, for $g_3'$ on a Hospital run. Thus, the observed correlations span most of the possible range of a Spearman coefficient.

The three measures also differ in how this variation arises. For $\mu^+$, the direction is unstable within datasets. On Adult, its correlation with recall is positive in exactly half of the 60 runs and negative in the other half. On Hospital, it is positive in 77\% of the 930 runs, but individual correlations range from $-0.61$ to $0.82$ with a standard deviation of $0.26$. The mean across Hospital runs, $0.1239$, therefore masks substantial run-to-run variation. On Claims, by contrast, the correlation is negative in all 165 runs.

The behavior of $\tau$ is different. Its correlation with recall is negative in 97\% of Adult runs and in every Hospital and Claims run. Rather than changing direction frequently, $\tau$ remains negative with respect to recall across nearly all runs. $g_3'$ shows a third pattern. Its correlation is positive in all Adult runs and in 93\% of Hospital runs, but negative in every Claims run. Section~\ref{sec:threats} shows that the negative direction of $g_3'$ on Claims persists under the alternative replacement strategies and the extended candidate space considered in our sensitivity analyses.

These results show that the relationship between AFD strength and recall depends strongly on both the strength measure and the dataset. Unlike the consistently positive association observed for precision, no common direction or magnitude emerges for recall across the three measures.

\subsubsection{RHS-Specific Heterogeneity on Hospital}
\label{sec:recall-attribute}

We next examine whether the variation observed for $\mu^+$ on Hospital is associated with particular RHS attributes. We focus on Hospital because its 48 minimal-FD targets span all 14 RHS attributes represented in the target pool, providing broad coverage across distinct RHS attributes. Adult covers only 2 distinct RHS attributes, and Claims covers only 3, with 6 of its 8 targets sharing the same RHS. These datasets therefore do not support a comparable analysis of variation across RHS attributes. A corresponding RHS-specific breakdown is not available for $\tau$ or $g_3'$. 

When the Hospital runs are grouped by RHS attribute, the relationship between $\mu^+$ and recall varies substantially. The mean correlation is positive for 9 of the 14 attributes. Provider Number shows the strongest positive association at $0.7743$, with Hospital Name, Zip Code, County Name, Phone Number, Hospital Type, Hospital Owner, Emergency Service, and StateAvg also positive. The correlation is negative for City, State, and Condition, with Condition showing the lowest mean correlation at $-0.4380$. Measure Code and Measure Name are both close to zero.

This distribution across RHS attributes shows that the broad variation observed across Hospital runs is not spread uniformly across the dataset. Instead, part of it is associated with systematic differences among RHS attributes: the same strength measure is positively associated with recall for some attributes, negatively associated with recall for others, and nearly unassociated for still others. Section~\ref{sec:mismatch} examines which structural properties may account for these differences.

\subsubsection{Extreme Individual Candidates}
\label{sec:recall-extremes}

Aggregate correlations can conceal extreme behavior of individual candidates. The two examples below are selected from such cases observed across the batch experiments rather than from a single run. They also illustrate why the RHS-specific averages reported above should not be interpreted as the behavior of every candidate sharing that RHS.

On Adult, consider
\[
\{\text{fnlwgt}, \text{education}, \text{hours-per-week}\} \to
\text{education-num}.
\]
This candidate has $\mu^+ = 0.9858$, among the highest strength scores observed on Adult, but recall of only $0.0061$, while its precision is $1.0$. Other Adult candidates combining very high $\mu^+$ with very low recall also frequently contain fnlwgt in their LHS.

The partition induced by the LHS of this candidate is highly fragmented. It contains $31{,}378$ equivalence classes, $93.11\%$ of candidate-valid rows lie in singleton classes, no non-singleton class contains more than $5$ rows, and $\mathrm{lhsUniqueness}=0.9637$. The high cardinality of fnlwgt contributes substantially to this fragmentation. As a result, most relevant errors occur in positions where the candidate has little opportunity to expose a within-class conflict, despite its near-maximal strength score.

A contrasting case appears on Hospital:
\[
\{\text{Condition}\} \to \text{Hospital Name}.
\]
This candidate has $\mu^+ = 0.0$, recall of $1.0$, and precision of $0.01$. Its LHS induces only $5$ equivalence classes, all of them non-singleton and together covering all $114{,}919$ rows. Within each class, the most frequent Hospital Name accounts for at most $1.27\%$ of the tuples, with most classes closer to $0.22\%$. The partition is therefore extremely coarse and the RHS values are highly heterogeneous within each class. Under the \textsc{Non-Majority} semantics, this structure makes relevant errors easy to flag, but also causes many unaltered cells to be flagged, yielding perfect recall together with very low precision.

The two examples illustrate opposite forms of mismatch. The Adult candidate combines near-maximal $\mu^+$ with almost no ability to recover relevant errors, despite perfect precision. The Hospital candidate combines zero $\mu^+$ with perfect recall but near-zero precision. These patterns are not confined to a single error rate or random seed. They recur across the experimental settings in which the candidates are evaluated.

Overall, the relationship between AFD strength and recall is not stable across measures, datasets, RHS attributes, or individual candidates. Section~\ref{sec:mismatch} examines structural factors that may account for part of this variation.

\subsection{Ranking Utility}
\label{sec:ranking-utility}

The preceding analyses examine whether AFD strength is associated with precision and recall. Candidate selection poses a more direct question: if candidates are ranked by strength, are the candidates placed near the top also those that perform well for error detection? We evaluate ranking utility through rank alignment, detection-cost regret, and agreement among strength measures on the top-ranked candidate.

\subsubsection{Rank Alignment}
\label{sec:ranking-topk}

We first examine whether candidates ranked near the top by strength are also ranked near the top by empirical precision or recall. For each run $u$, we measure this using top-$k$ overlap,
\[
O_k^{(u)}
=
\frac{
	\left|
	T_k^{S}(u)\cap T_k^{Y}(u)
	\right|
}{k},
\]
where $T_k^{S}(u)$ and $T_k^{Y}(u)$ denote the top-$k$ candidate sets ranked by strength and by detection metric $Y$, respectively. Table~\ref{tab:topk-overlap} reports the mean overlap across runs for $k=10$.

\begin{table}[t]
	\centering
	\caption{Mean top-10 overlap between rankings by AFD strength and rankings by empirical precision and recall, by measure and dataset. The final column pools all 1,155 runs across datasets and corruption mechanisms. Results for $k=1$ and $k=5$ show the same overall pattern and are omitted.}
	\label{tab:topk-overlap}
	\begin{subtable}{0.48\textwidth}
		\centering
		\caption{Overlap against precision}
		\label{tab:topk-overlap-precision}
		\begin{tabular}{lcccc}
			\toprule
			Measure & Adult & Hospital & Claims & Pooled \\
			\midrule
			$\mu^+$   & 0.045 & 0.004 & 0.000 & 0.006 \\
			$\tau$    & 0.000 & 0.000 & 0.000 & 0.000 \\
			$g_3'$    & 0.462 & 0.016 & 0.000 & 0.037 \\
			\bottomrule
		\end{tabular}
	\end{subtable}
	
	\hfill
	
	\begin{subtable}{0.48\textwidth}
		\centering
		\caption{Overlap against recall}
		\label{tab:topk-overlap-recall}
		\begin{tabular}{lcccc}
			\toprule
			Measure & Adult & Hospital & Claims & Pooled \\
			\midrule
			$\mu^+$   & 0.010 & 0.002 & 0.000 & 0.002 \\
			$\tau$    & 0.000 & 0.000 & 0.000 & 0.000 \\
			$g_3'$    & 0.587 & 0.014 & 0.000 & 0.042 \\
			\bottomrule
		\end{tabular}
	\end{subtable}
\end{table}

The overlap is close to zero in most measure-dataset combinations. For $\tau$, the top-10 overlap is exactly zero for both precision and recall on all three datasets. The values for $\mu^+$ are also very small: the largest is $0.045$ against precision on Adult, while all values on Hospital and Claims are at or below $0.004$. On Claims, none of the three strength measures shares any top-10 candidate with either the precision ranking or the recall ranking.

The main exception is $g_3'$ on Adult. Its top-10 overlap reaches $0.462$ against precision and $0.587$ against recall. This pattern does not carry over to the other datasets. On Hospital, the corresponding values fall to $0.016$ and $0.014$, and on Claims both are zero.

These results show that a candidate ranked highly by AFD strength is generally not among the candidates ranked highly by empirical detection performance. The exception is $g_3'$ on Adult, where the strength ranking shows substantial overlap with both precision and recall rankings. The next analysis asks whether these ranking differences also translate into a cost when strength is used to choose a single candidate.

\subsubsection{Detection-Cost Regret}
\label{sec:ranking-regret}

As defined in Section~\ref{sec:prelim-ldetect}, regret is the additional detection cost incurred by selecting the top-ranked candidate under a strength measure instead of the candidate with the minimum $L_{detect}$. A regret of zero therefore means that the selected candidate has the minimum detection cost.

\begin{table}[t]
	\centering
	\caption{Mean regret of the top-1 candidate selected by each measure, by dataset, pooled across both corruption mechanisms. The final column of each subtable pools all runs across datasets and mechanisms. Normalized regret divides absolute regret by the minimum detection cost, floored at $1$, as defined in Section~\ref{sec:prelim-ldetect}.}
	\label{tab:regret-both}
	\begin{subtable}{0.48\textwidth}
		\centering
		\caption{Absolute regret}
		\label{tab:regret}
		\begin{tabular}{lcccc}
			\toprule
			Measure & Adult & Hospital & Claims & Pooled \\
			\midrule
			$\mu^+$   & 1619.22 & 6127.66 & 4398.45 & 5646.43 \\
			$g_3'$    & 7.77    & 6127.66 & 4398.45 & 5562.71 \\
			$\tau$    & 1655.62 & 6127.66 & 4398.45 & 5648.32 \\
			\bottomrule
		\end{tabular}
	\end{subtable}
	
	\hfill
	
	\begin{subtable}{0.48\textwidth}
		\centering
		\caption{Normalized regret}
		\label{tab:regret-normalized}
		\begin{tabular}{lcccc}
			\toprule
			Measure & Adult & Hospital & Claims & Pooled \\
			\midrule
			$\mu^+$   & 1619.22 & 5669.93 & 1060.24 & 4800.98 \\
			$g_3'$    & 7.77    & 5669.93 & 1060.24 & 4717.27 \\
			$\tau$    & 1655.62 & 5669.93 & 1060.24 & 4802.87 \\
			\bottomrule
		\end{tabular}
	\end{subtable}
\end{table}

Table~\ref{tab:regret} reports mean absolute regret. On Hospital, all three strength measures incur the same mean regret, $6127.66$. The same occurs on Claims, where the mean is $4398.45$ for every measure. As shown in Section~\ref{sec:ranking-agreement}, this equality arises because the three measures select the same top-1 candidate in every run on these two datasets. The shared choice does not have the minimum detection cost, so all three measures incur the same nonzero regret.

Adult behaves differently. The mean absolute regret is $1619.22$ for $\mu^+$ and $1655.62$ for $\tau$, but only $7.77$ for $g_3'$. This small mean for $g_3'$ does not reflect uniformly small regret in every run. Instead, $g_3'$ selects a candidate with the minimum detection cost in $31$ of the $60$ Adult runs, giving a median regret of $0$. A small number of runs with much larger regret account for most of the remaining mean. By contrast, $\mu^+$ and $\tau$ do not select a candidate with the minimum detection cost in any of the 60 Adult runs, and both have median regret above $1500$.

Table~\ref{tab:regret-normalized} adjusts these values for differences in the minimum detection cost. On Adult, normalization has almost no effect because the minimum detection cost is usually at or near the denominator floor of $1$. On Hospital, the mean regret decreases only modestly, from $6127.66$ to $5669.93$, because the minimum detection cost is zero in most runs and reaches at most $25$. On Claims, the effect is much larger: the mean regret falls from $4398.45$ to $1060.24$. The minimum detection cost on Claims is typically nonzero, with a median of $4$ and a maximum of $19$, so the same absolute excess cost is smaller relative to the minimum detection cost.

Absolute and normalized regret therefore lead to the same qualitative conclusion. On Hospital and Claims, the candidate selected by all three strength measures consistently has a higher detection cost than the minimum-cost candidate. On Adult, $g_3'$ is a clear exception, selecting a candidate with the minimum detection cost in just over half of the runs, whereas $\mu^+$ and $\tau$ do not do so in any run. These results show that high strength does not by itself provide a reliable basis for choosing the candidate that minimizes detection cost.

\subsubsection{Top-1 Agreement}
\label{sec:ranking-agreement}

This part examines top-1 agreement among $\mu^+$, $\tau$, and $g_3'$. We first ask: how often do the three measures select the same top-ranked candidate? We then analyze the structural patterns of these shared selections and whether the selected candidate has the minimum detection cost.

The answer differs sharply across datasets. On Hospital, $\mu^+$, $\tau$, and $g_3'$ select the same top-1 candidate in all $930$ runs. On Claims, they also agree in every run: all $120$ FD-aligned runs and all $45$ general-corruption runs, for a total of $165$. Adult shows almost no such three-way agreement. Despite the complete agreement on Hospital and Claims, the shared top-1 candidate never has the minimum detection cost on either dataset. Table~\ref{tab:agreement-structure} summarizes these patterns.

\begin{table*}[t]
	\centering
	\caption{Top-1 agreement among strength measures and the structural
		patterns associated with their selected candidates. Agreement is
		the fraction of runs in which $\mu^+$, $\tau$, and $g_3'$ select
		the same top-1 candidate. Detection-cost optimality reports how
		often the candidate selected by each strength measure minimizes
		$L_{detect}$.}
	\label{tab:agreement-structure}
	\renewcommand{\arraystretch}{1.5}
	\begin{tabular}{lp{2.6cm}p{4.6cm}p{3cm}p{3.8cm}}
		\toprule
		Dataset & Agreement among measures & Dominant LHS pattern &
		Structural concentration & Detection-cost optimality \\
		\midrule
		Hospital &
		$930/930$ ($100\%$) &
		$\{\text{Provider Number}, \text{Measure Code}\}$, exact key on
		valid rows; structurally similar substitute when RHS is Provider
		Number &
		$870/930$ runs share primary LHS &
		$0\%$ (all measures) \\
		Claims &
		$165/165$ ($100\%$) &
		$\{\text{Claim Number}\}$, exact key &
		$165/165$ runs share same LHS &
		$0\%$ (all measures) \\
		Adult &
		$1/60$ ($1.7\%$) &
		no dominant LHS; fnlwgt recurs among winners &
		$14/18$ distinct winners contain fnlwgt &
		$0\%$ ($\mu^+$, $\tau$); $31/60$ ($51.7\%$, $g_3'$) \\
		\bottomrule
	\end{tabular}
\end{table*}

On Hospital, the three measures agree in every run, but the selected candidate changes with the RHS attribute under evaluation. In $870$ of the $930$ runs, spanning $12$ RHS attributes, the selected candidate has the LHS
\[
\{\text{Provider Number}, \text{Measure Code}\}.
\]
This LHS induces an extreme partition structure. For $12$ of the $13$ RHS attributes with which it forms a valid candidate, there are $114{,}919$ candidate-valid rows and exactly $114{,}919$ equivalence classes. Every class is therefore a singleton, and $\mathrm{lhsUniqueness}=1$. County Name is the only exception in row count: it has $114{,}869$ candidate-valid rows because $50$ rows contain a null in that RHS attribute, but the same one-row-per-class pattern remains.

The remaining $60$ Hospital runs have Provider Number as the RHS, so $\{\text{Provider Number}, \text{Measure Code}\}$ cannot be used as the LHS of a nontrivial candidate. The selected candidate instead uses
\[
\{\text{Hospital Name}, \text{City}, \text{StateAvg}\}
\]
as its LHS. This alternative also has a near-key partition structure. Thus, the complete agreement among the three strength measures on Hospital is associated with repeatedly selecting candidates whose LHSs partition the valid rows into singleton or near-singleton classes.

Claims shows an even simpler pattern. Across all $165$ runs, the selected candidate always has the LHS
\[
\{\text{Claim Number}\}.
\]
The RHS varies among Airport Code, Airport Name, and Disposition. For the candidates with RHS Airport Code and Airport Name, the candidate-valid subrelation contains $89{,}207$ rows in each case. For the candidate with RHS Disposition, it contains $87{,}097$ rows. In all three cases, the number of equivalence classes equals the number of candidate-valid rows, so $\mathrm{lhsUniqueness}=1$. Claim Number therefore uniquely identifies every candidate-valid tuple.

None of the targets in the Claims pool has Claim Number as its RHS. The LHS $\{\text{Claim Number}\}$ is therefore available for every target considered in the batch experiments. Because both corruption mechanisms modify only RHS values, this key-like LHS remains unchanged across runs. The corresponding candidates retain maximal strength and are repeatedly selected by all three measures. However, as shown by the regret results above, these shared selections never have the minimum detection cost.

Adult provides the opposite pattern. The $60$ runs produce $18$ distinct winning candidates, and no single candidate is selected in more than $10$ runs. Fourteen of the $18$ winners contain fnlwgt in their LHS, so high-cardinality structure still appears frequently, but no single LHS dominates the candidate space as it does on Hospital and Claims. Three-way agreement among $\mu^+$, $\tau$, and $g_3'$ occurs in only $1$ of the $60$ runs. Pairwise agreement is also limited: $\mu^+$ and $\tau$ agree in $43\%$ of FD-aligned runs and $33\%$ of general-corruption runs, while $\mu^+$ and $g_3'$ agree in at most $3\%$ of runs and never agree under general corruption.

Across the three datasets, agreement among strength measures therefore varies with how strongly the top-ranked candidates concentrate around recurring key-like LHS structures. Hospital and Claims contain such dominant patterns and show complete three-way agreement, whereas Adult does not and shows little agreement.

The three ranking analyses lead to the same conclusion. Rankings by strength usually have little overlap with rankings by empirical detection performance, and selecting the top-ranked candidate by strength can incur substantial detection cost. Even agreement among different strength measures does not imply that the selected candidate has the minimum detection cost.

\section{Understanding the Strength-Value Mismatch}
\label{sec:mismatch}

Section~\ref{sec:main-results} shows that AFD strength does not consistently align with detection value. We now examine why. We consider three possible explanations: differences between FD-aligned and general corruption, the equivalence-class structure defined by the LHS of a candidate, and other structural factors that may contribute to the remaining variation.

\subsection{Corruption Mechanism}
\label{sec:mismatch-mechanism}

We first ask whether the mismatch can be explained by how errors are injected. If FD-aligned and general corruption produced substantially different relationships between strength and detection performance, the corruption mechanism would be a plausible source of the instability observed in Section~\ref{sec:main-results}.

The largest differences between the two corruption mechanisms appear on Adult. For $\tau$ and $g_3'$, the mean strength-precision correlation differs by roughly $0.1$ between FD-aligned and general corruption. No comparable difference appears on Hospital or Claims. The strength-recall correlations show the same broad pattern: the difference between the two mechanisms is more visible on Adult and much smaller on Hospital and Claims.

The same pattern across datasets also appears for the relationship between structural detection opportunity and recall. Section~\ref{sec:mismatch-geometry} defines the non-singleton row fraction $\rho_{ns}$, which measures the fraction of candidate-valid rows that belong to non-singleton equivalence classes. We compute its mean run-level Spearman correlation with recall separately under FD-aligned and general corruption. The absolute difference between the two correlations is $0.119$ on Adult, $0.060$ on Hospital, and $0.005$ on Claims. Thus, across both strength-based and structure-based comparisons, changing the corruption mechanism has its largest effect on Adult and its smallest effect on Claims.

These differences show that corruption mechanism does affect the observed relationships. However, the effect is limited and strongly dataset-dependent. In particular, it is too small to account for the much larger variation in strength-recall correlation observed across measures and datasets, where individual-run correlations range from $-0.8267$ to $0.8867$. Corruption mechanism therefore cannot serve as the primary explanation for the strength-value mismatch.

One possible reason for the weaker mechanism effect on Hospital and Claims is the prevalence of key-like LHSs in their candidate spaces. As discussed in Section~\ref{sec:mismatch-geometry}, some candidates on these datasets have LHSs that uniquely or almost uniquely identify valid tuples. Because both corruption mechanisms modify only RHS values, these LHS partitions remain unchanged across runs. Rankings dominated by such candidates may therefore be less sensitive to how errors are injected. We treat this as a possible explanation rather than a verified result.

\subsection{Structural Detection Opportunity}
\label{sec:mismatch-geometry}

A candidate can have high AFD strength even when its LHS partition leaves little opportunity for an injected RHS error to produce an observable conflict. We therefore examine the structural opportunity for error detection provided by the LHS partition of the candidate.

\subsubsection{Structural Opportunity from the LHS Partition}

For a candidate $\phi:X\to A$, consider the candidate-specific equivalence classes obtained by restricting the LHS partition $\Pi_X$ to $r_\phi$, as defined in Section~\ref{sec:prelim-afd}. An RHS error can produce a within-class conflict only if its row belongs to a class containing at least one other tuple. This motivates the \emph{non-singleton row fraction}
\[
\rho_{ns}(\phi)
=
\frac{\sum_{c:\,|c|>1}|c|}{N_\phi},
\]
where the sum is over the candidate-specific equivalence classes of $\phi$.

Thus, $\rho_{ns}(\phi)$ is the fraction of candidate-valid rows that belong to non-singleton equivalence classes. A row in a singleton class has no other tuple with the same LHS value, so changing its RHS value cannot by itself create a conflict within that class. $\rho_{ns}(\phi)$ therefore measures how much structural opportunity candidate $\phi$ provides for an RHS error to become detectable through a within-class disagreement.

$\rho_{ns}(\phi)$ is determined entirely by the LHS partition of the candidate and does not vary across runs for the same candidate. This distinguishes it from the visibility ratio introduced in Section~\ref{sec:prelim-vuse},
\[
\rho_{vis}(\phi)
=
\frac{|E_{vis}(\phi)|}{|E_{rel}(\phi)|},
\]
which measures, for a particular run, the fraction of relevant injected errors whose rows belong to conflicting classes after corruption. Thus, $\rho_{ns}$ describes structural opportunity from the LHS partition, whereas $\rho_{vis}$ describes realized structural visibility after corruption.

In our experiments, the mean values of $\rho_{ns}$ and $\rho_{vis}$ differ by less than two percentage points for every dataset and corruption mechanism. We therefore use $\rho_{ns}$ as the primary structural quantity in the analyses below and retain $\rho_{vis}$ as a check based on the errors actually injected in each run.

\subsubsection{Tracking Recall Through Structural Opportunity}

We next compare how strongly AFD strength and structural opportunity are associated with recall. Table~\ref{tab:conflict-geometry} reports the mean run-level Spearman correlation with recall for the three strength measures, $\rho_{ns}$, and $\rho_{vis}$.

\begin{table}[t]
	\centering
	\caption{Mean run-level Spearman correlation with recall for the
		three AFD strength measures and the two structural quantities
		$\rho_{ns}$ and $\rho_{vis}$. The final column pools all
		1,155 runs across datasets and corruption mechanisms.}
	\label{tab:conflict-geometry}
	\begin{tabular}{lcccc}
		\toprule
		Measure & Adult & Hospital & Claims & Pooled \\
		\midrule
		$\mu^+$       & 0.0129  & 0.1239  & -0.4869 & 0.0309 \\
		$g_3'$        & 0.5688  & 0.2669  & -0.3756 & 0.1908 \\
		$\tau$        & -0.2285 & -0.5056 & -0.7312 & -0.5234 \\
		$\rho_{ns}$   & 0.7387  & 0.8521  & 0.9472  & 0.8598 \\
		$\rho_{vis}$  & 0.7418  & 0.8623  & 0.9540  & 0.8691 \\
		\bottomrule
	\end{tabular}
\end{table}

The difference between the strength measures and the two structural quantities is clear. The correlation of $\rho_{ns}$ with recall is strongly positive on all three datasets, ranging from $0.7387$ on Adult to $0.9472$ on Claims. $\rho_{vis}$ shows an almost identical pattern, with correlations between $0.7418$ and $0.9540$. Neither structural quantity changes direction across datasets.

The three strength measures behave differently. $\mu^+$ is close to zero on Adult, weakly positive on Hospital, and negative on Claims. $g_3'$ is positive on Adult and Hospital but negative on Claims. $\tau$ is negative on all three datasets. Thus, although the three strength measures fail to track recall in different ways, none shows the consistently strong positive relationship observed for $\rho_{ns}$ and $\rho_{vis}$. The structure of the LHS partition is therefore much more closely associated with whether relevant errors are recovered than AFD strength itself.

For a fixed candidate $\phi$, both $r_\phi$ and its candidate-specific LHS partition remain unchanged across runs, because the corruption mechanisms modify only non-null RHS values. Therefore, $\rho_{ns}(\phi)$ is fixed for that candidate and does not depend on which errors are injected in a particular run. This is also consistent with the experimental results: across all $8{,}398$ distinct candidates in the batch experiments, the numbers of equivalence classes and non-singleton classes remain identical in every run in which the same candidate appears.

\subsubsection{A Structural Bound on Recall}

The strong association between structural opportunity and recall has a direct explanation under the \textsc{Non-Majority} detection semantics. 

For a particular corruption-and-detection run $u$, let
$E_{rel}^{(u)}(\phi)$, $E_{vis}^{(u)}(\phi)$, and $TP_{\phi}^{(u)}$
denote the run-specific relevant-error set, visible-error set, and true-positive set for candidate $\phi$, respectively. Define
\[
\rho_{vis}^{(u)}(\phi)
=
\frac{|E_{vis}^{(u)}(\phi)|}
{|E_{rel}^{(u)}(\phi)|}.
\]

Let $E_{ns}^{(u)}(\phi)$ denote the relevant injected errors whose rows belong to non-singleton candidate-specific equivalence classes of $\phi$. Define
\[
q_{ns}^{(u)}(\phi)
=
\frac{|E_{ns}^{(u)}(\phi)|}
{|E_{rel}^{(u)}(\phi)|}.
\]
Thus, $q_{ns}^{(u)}(\phi)$ is the fraction of relevant errors injected in run $u$ that fall on rows where a within-class conflict is structurally possible.

Unlike $\rho_{ns}(\phi)$, which is computed over all candidate-valid rows, $q_{ns}^{(u)}(\phi)$ is computed only over the relevant errors actually injected in that run. It also differs from $\rho_{vis}^{(u)}(\phi)$. An error contributes to $q_{ns}^{(u)}(\phi)$ if its row belongs to a non-singleton class, regardless of whether that class is conflicting after corruption. It contributes to $\rho_{vis}^{(u)}(\phi)$ only if the class is conflicting after corruption.

Under \textsc{Non-Majority}, a true positive must arise from a relevant error whose row belongs to a conflicting class after corruption. Every conflicting class is non-singleton. Therefore,
\[
TP_{\phi}^{(u)}
\subseteq
E_{vis}^{(u)}(\phi)
\subseteq
E_{ns}^{(u)}(\phi),
\]
and hence
\[
\mathrm{recall}_{\phi}^{(u)}
\le
\rho_{vis}^{(u)}(\phi)
\le
q_{ns}^{(u)}(\phi).
\]
This is a strict run-specific relationship. It follows directly from the detection semantics and does not require any assumption about how the corrupted positions are sampled.

Suppose that, conditional on injecting $m$ relevant errors for candidate $\phi$, the corrupted positions are sampled uniformly from its candidate-valid RHS positions. Then
\[
\mathbb{E}\!\left[
q_{ns}^{(u)}(\phi)
\mid
|E_{rel}^{(u)}(\phi)|=m
\right]
=
\rho_{ns}(\phi).
\]
Under this sampling model, $\rho_{ns}(\phi)$ is the expected fraction of relevant injected errors that fall on non-singleton rows.

Our experiments do not guarantee this sampling condition for every candidate. In particular, FD-aligned corruption selects positions using the LHS partition of the target FD, whereas the candidates evaluated for that target may have different LHS partitions. These partitions may therefore be statistically related. We consequently do not treat $\rho_{ns}(\phi)$ as a strict bound on recall in an individual run.

The per-class corruption cap does not appear to have caused substantial truncation at the aggregate level. All $1{,}155$ runs achieve the requested number of corruptions, and the requested count never exceeds $10\%$ of the eligible cells, compared with the $50\%$ cap applied within any one eligible class. These observations do not establish uniform sampling across candidate partitions, but they rule out failure to reach the requested corruption count as an explanation for the results below.

Empirically, however, the static $\rho_{ns}$ provides a very close upper reference for recall. Across all $513{,}480$ candidate-run records, recall exceeds $\rho_{ns}$ in only $0.58\%$ of cases, corresponding to $2{,}995$ records. No record on any dataset exceeds $\rho_{ns}+0.05$.

At the same time, recall usually remains well below the available structural opportunity. For each candidate-run record, we compute
\[
\mathrm{recall}_{\phi}^{(u)}-\rho_{ns}(\phi)
\]
and then average this difference within each dataset. The mean is $-0.1468$ on Adult, $-0.2058$ on Hospital, and $-0.1687$ on Claims. Thus, $\rho_{ns}$ captures a necessary opportunity for detection, but recall can remain substantially lower than $\rho_{ns}$.

The small fraction of records in which recall exceeds $\rho_{ns}$ is concentrated near the upper end of the range. Among the $2{,}995$ such records, $95.1\%$ have $\rho_{ns}>0.9$, and their median $\rho_{ns}$ is $0.9998$. The amount by which recall exceeds $\rho_{ns}$ is also very small: the median excess is $0.0001$, and $97.8\%$ of the excesses are below $0.01$. These observations show that the few cases in which recall exceeds $\rho_{ns}$ involve only small excesses and occur mainly when $\rho_{ns}$ is already close to one.

The distinction is therefore important. The static quantity $\rho_{ns}(\phi)$ is not a strict run-specific bound. Empirically, however, it provides a close upper reference for recall in our experiments.

\subsubsection{Robustness Across Datasets}

We next examine whether the relationship between $\rho_{ns}$ and recall persists under different data distributions and across different RHS attributes. On Hospital, we compare the relationship separately for the 14 RHS attributes in the target pool. On Claims and Adult, we examine how $\rho_{ns}$ is distributed across all distinct candidates and whether the correlation with recall remains strong under very different distributions. Hospital is analyzed by RHS attribute because its target pool spans all 14 RHS attributes, whereas the Adult and Claims target pools cover only 2 and 3 distinct RHS attributes, respectively.

For the RHS-specific analysis on Hospital, we compare $\rho_{ns}$ with $\mu^+$ only, because the per-attribute breakdown reported in Section~\ref{sec:recall-attribute} is available for $\mu^+$ but not for $\tau$ or $g_3'$.

\begin{figure}[t]
	\centering
	\includegraphics[width=\linewidth]{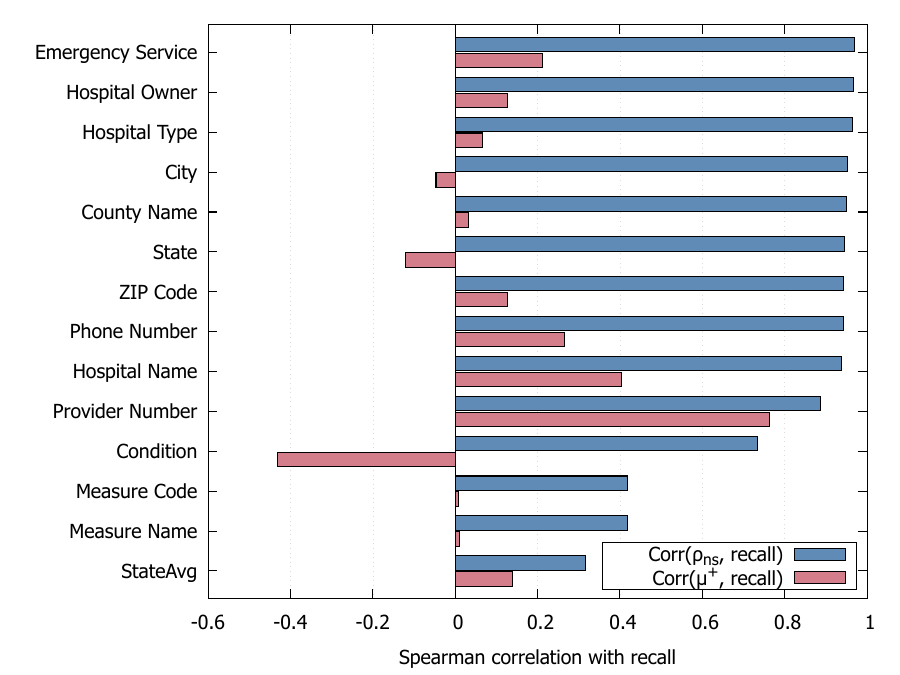}
	\caption{Correlation with recall for $\mu^+$ and $\rho_{ns}$,
		by RHS attribute on Hospital, sorted by $\rho_{ns}$.}
	\label{fig:hospital-rhs-heterogeneity}
\end{figure}

Figure~\ref{fig:hospital-rhs-heterogeneity} compares the correlation of recall with $\mu^+$ and with $\rho_{ns}$ for the 14 RHS attributes in Hospital. As reported in Section~\ref{sec:recall-attribute}, the correlation between $\mu^+$ and recall is positive for 9 attributes, negative for 3, and close to zero for 2. In contrast, the correlation between $\rho_{ns}$ and recall is positive for all 14 attributes, ranging from $0.32$ for StateAvg to $0.97$ for Emergency Service.

The contrast is especially clear for Condition. For this RHS attribute, the correlation with recall is $-0.4380$ for $\mu^+$ but $0.73$ for $\rho_{ns}$. Thus, $\rho_{ns}$ remains positively associated with recall even for RHS attributes on which $\mu^+$ is negatively associated with recall. This does not explain why the correlation of $\mu^+$ with recall changes sign across RHS attributes. It shows only that $\rho_{ns}$ captures a different relationship with recall. We return to the remaining RHS-specific variation in Section~\ref{sec:mismatch-open}.

\begin{figure}[t]
	\centering
	\includegraphics[width=\linewidth]{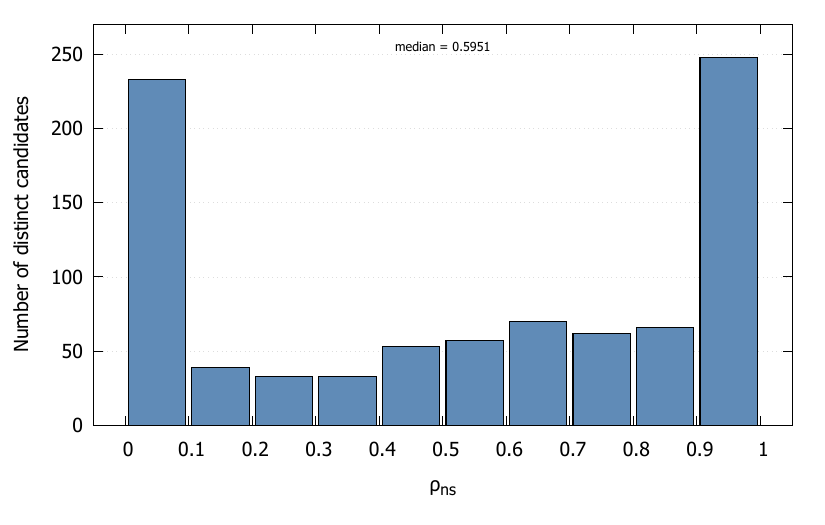}
	\caption{Distribution of $\rho_{ns}$ across the 894 distinct
		candidates in the candidate space of Claims.}
	\label{fig:claims-rho-ns-dist}
\end{figure}

The strong correlation on Claims is not confined to candidates with similar values of $\rho_{ns}$. Figure~\ref{fig:claims-rho-ns-dist} shows the distribution of $\rho_{ns}$ across the 894 distinct candidates in the Claims candidate space. The distribution spans almost the entire interval from $0$ to $1$ and is strongly U-shaped. A total of $233$ candidates fall in $[0,0.1)$ and $248$ fall in $[0.9,1.0)$. Every intermediate decile still contains between $33$ and $70$ candidates. The median is $0.5951$, and the interquartile range extends from $0.0591$ to $0.9522$.

The pooled correlation between $\rho_{ns}$ and recall on Claims is $0.9472$. This strong correlation therefore holds across candidates with widely different amounts of structural opportunity, rather than within a narrow range of $\rho_{ns}$.

\begin{figure}[t]
	\centering
	\includegraphics[width=\linewidth]{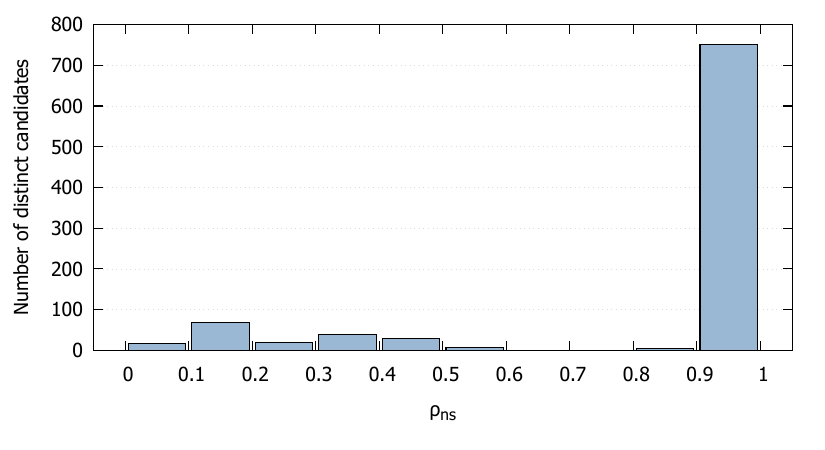}
	\caption{Distribution of $\rho_{ns}$ across the 938 distinct
		candidates in the candidate space of Adult.}
	\label{fig:adult-rho-ns-dist}
\end{figure}

Adult has a very different distribution. Figure~\ref{fig:adult-rho-ns-dist} shows that $80\%$ of its 938 distinct candidates have $\rho_{ns}>0.9$, while no candidate falls in the interval $[0.6,0.8)$. Thus, most Adult candidates have $\rho_{ns}$ close to $1$, rather than spanning the full range observed on Claims. Even under this much narrower distribution, the pooled correlation between $\rho_{ns}$ and recall remains positive at $0.7387$.

These results show that the relationship between structural opportunity and recall is not limited to one dataset or one particular distribution of $\rho_{ns}$. It remains positive across all 14 RHS attributes examined on Hospital, across the broad range of $\rho_{ns}$ values on Claims, and across the much more concentrated distribution on Adult.

\subsubsection{From Visibility to Detection}

Structural visibility does not guarantee detection under the \textsc{Non-Majority} semantics. A relevant error may belong to a conflicting class and therefore be included in $E_{vis}$, but the detector can still miss it. In particular, \textsc{Non-Majority} produces no flags for a conflicting class when more than one RHS value shares the highest frequency.

Under \textsc{Non-Majority}, every true positive belongs to $E_{vis}^{(u)}(\phi)$. Therefore, for a run $u$ with $|E_{vis}^{(u)}(\phi)|>0$, the visible recall can be written as
\[
\mathrm{recall}_{vis,\phi}^{(u)}
=
\frac{|TP_{\phi}^{(u)}|}
{|E_{vis}^{(u)}(\phi)|}.
\]
Since
\[
\rho_{vis}^{(u)}(\phi)
=
\frac{|E_{vis}^{(u)}(\phi)|}
{|E_{rel}^{(u)}(\phi)|},
\]
the default recall can be written as
\[
\mathrm{recall}_{\phi}^{(u)}
=
\rho_{vis}^{(u)}(\phi)
\cdot
\mathrm{recall}_{vis,\phi}^{(u)}.
\]
Thus, $\rho_{vis}$ measures how much of the relevant error set becomes structurally visible, while $\mathrm{recall}_{vis}$ measures how much of the visible error set is actually detected.

We examine whether majority ties are associated with the second stage. Across records with nonzero visible ground truth, the Spearman correlation between the number of conflicting classes with no unique majority and $\mathrm{recall}_{vis}$ is $-0.7322$ on Adult, $-0.7209$ on Hospital, and $-0.7135$ on Claims. The direction and magnitude are therefore highly consistent across all three datasets: more conflicting classes with no unique majority are associated with lower recall among errors that are already structurally visible.

Using the same candidate-run records with nonzero visible ground truth, pooled across both corruption mechanisms, we compare records at or below the dataset-specific 10th percentile of $\mathrm{recall}_{vis}$ with those at or above the 90th percentile. Ties at the percentile thresholds are retained. The same broad pattern appears. On Hospital, the median number of conflicting classes with no unique majority is 0 in the high-$\mathrm{recall}_{vis}$ group and 2,978 in the low-$\mathrm{recall}_{vis}$ group. On Adult, the corresponding medians are 2 and 1,560, and on Claims they are 10 and 2,637. The fraction of records containing at least one conflicting class with no unique majority also increases sharply in the low-$\mathrm{recall}_{vis}$ groups. The contrast is especially clear on Adult and Hospital, while many Claims records already contain such classes even when $\mathrm{recall}_{vis}$ is high.

These results identify majority ties as an important source of missed detections after structural visibility has been established. Non-singleton structure determines whether an error has an opportunity to participate in a conflict, and $\rho_{vis}$ measures how much of the relevant error set is structurally visible after corruption. Under \textsc{Non-Majority}, detection can still fail when the conflicting class has no unique majority. Structural opportunity, structural visibility, and actual detection therefore represent distinct stages of the detection process.

\subsection{Open Questions}
\label{sec:mismatch-open}

Sections~\ref{sec:mismatch-mechanism} and~\ref{sec:mismatch-geometry} explain part of the observed strength-value mismatch, but two questions remain unresolved.

The first concerns the sign reversal in the correlation between $\mu^+$ and recall across RHS attributes on Hospital. As reported in Section~\ref{sec:recall}, the correlation is positive for $9$ of the $14$ RHS attributes, negative for $3$, namely City, State, and Condition, and close to zero for the remaining $2$. Structural detection opportunity does not explain this split because the correlation between $\rho_{ns}$ and recall is positive for all $14$ attributes. We test two additional explanations for this sign reversal.

\begin{itemize}
	\item The first explanation concerns RHS cardinality. Hospital Type has $4$ distinct values and Emergency Service has $2$, yet both show positive $\mu^+$--recall correlation. Condition has $5$ distinct values but shows negative correlation. The three negative attributes also span a wide range of cardinalities: Condition has $5$ distinct values, State has $55$, and City has $2{,}727$. RHS cardinality therefore does not distinguish the positive and negative cases.
	
	\item The second explanation concerns how often an RHS attribute appears in the LHS of other dependencies. Condition never appears as an LHS attribute in the other dependencies considered here, but the same is true of Hospital Type and Emergency Service, even though their $\mu^+$--recall correlations are positive. These counterexamples do not show that RHS properties are irrelevant. They show only that the two RHS properties examined here are insufficient to explain why the correlation changes sign across attributes.
\end{itemize}

The second unresolved question concerns candidates with very high LHS uniqueness. Section~\ref{sec:ranking-utility} shows that exact-key or near-key LHSs can create severe ranking failures. When the LHS partition contains only singleton classes, $\mu^+$, $\tau$, and $g_3'$ reach their maximum values even though the candidate provides no within-class opportunity for detecting RHS errors.

A natural hypothesis is therefore that candidates with unusually high LHS uniqueness may be especially prone to strength-value mismatch. Hospital and Claims support this hypothesis: candidates with the largest mismatch tend to have high LHS uniqueness. Adult does not show the same pattern, and this difference persists across corruption mechanisms and random seeds. High LHS uniqueness can therefore produce severe failure modes, but it is not a necessary condition for strength-value mismatch.

These analyses show that several simple explanations are insufficient. RHS cardinality, frequency of RHS attributes in other LHSs, and high LHS uniqueness each explain some cases but do not account for the change in strength-recall correlation across datasets and RHS attributes.

\section{Discussion}
\label{sec:discussion}

This section returns to the three questions posed in the Introduction. AFD strength is positively associated with precision across all three datasets, although the association is weaker for $\tau$ than for $\mu^+$ and $g_3'$. Its relationship with recall is much less stable and varies substantially across measures and datasets.

The mismatch also appears in candidate ranking. Candidates ranked highly by strength often do not rank highly by observed detection performance. When different strength measures select the same top candidate, that candidate is often a key-like or near-key case and is not the detection-cost optimum.

The structural analysis helps explain this difference. Recall depends on whether an injected error occurs in a row where the LHS partition of the candidate allows a within-class conflict to arise. The non-singleton row fraction $\rho_{ns}$ measures this structural opportunity and is strongly positively associated with recall on all three datasets. For a particular run, recall is strictly bounded by the fraction of relevant injected errors that fall on non-singleton rows. The static quantity $\rho_{ns}$ closely approximates this structural limit in our experiments. This relationship is much more stable than the relationship between recall and any of the three AFD strength measures.

Structural visibility is still not sufficient for detection. Under the \textsc{Non-Majority} semantics, a conflicting class can produce no flags when the largest RHS frequency is tied. The analysis in Section~\ref{sec:mismatch-geometry} shows that more tied conflicting classes are consistently associated with lower recall among errors that are already structurally visible. Thus, the LHS partition determines structural opportunity, post-corruption conflicts determine visibility, and majority ties can further reduce detection after visibility has been established.

These results also clarify why precision and recall behave differently. Precision is evaluated only after cells have been flagged. Recall, by contrast, also depends on whether relevant errors occur in structurally detectable locations. The results suggest that precision and recall depend on different properties of the candidate and the LHS partition. AFD strength is associated with the former more consistently than with the latter.

Several factors explain part of the mismatch, but none explains all of it. Corruption mechanism affects the observed relationships most strongly on Adult and much less on Hospital and Claims. Structural detection opportunity explains a large part of the recall variation, but it does not explain why the correlation between $\mu^+$ and recall changes sign across RHS attributes on Hospital. We also test two RHS properties: attribute cardinality and how often the RHS attribute appears in the LHS of other dependencies. Neither explains this sign reversal. High LHS uniqueness creates severe failure modes on Hospital and Claims, but Adult shows that high uniqueness is not necessary for strength-value mismatch.

We can therefore explain several important failure modes, but not why the strength-recall correlation is positive in some cases and negative in others. The remaining variation indicates that additional structural or data-dependent factors are still missing from the current analysis.

The broader implication is that AFD strength and downstream detection value should be evaluated separately. A measure may quantify approximate dependency satisfaction well and still align poorly with recall-oriented error detection. Stronger satisfaction does not by itself imply that the errors violating a dependency will be easier to find. This is the sense in which strength is not value.

\section{Threats to Validity}
\label{sec:threats}

We evaluate five potential threats to the robustness and scope of our findings: replacement strategy, candidate-space depth, false-negative versus false-positive weighting, detection semantics, and dataset choice. The main qualitative conclusions remain stable under these checks, although the relative performance of individual measures changes under some experimental settings.

\subsection{Replacement Strategy Sensitivity}
\label{sec:threats-strategy}

The batch experiments use \texttt{CROSS\_CLASS} for FD-aligned corruption and \texttt{RANDOM\_COPY} for general corruption. To check whether the findings depend on these replacement strategies, we conduct a separate batch of $192$ runs using the two alternatives, \texttt{WITHIN\_DOMAIN} and \texttt{FRESH\_VALUE}, under both corruption mechanisms, two error rates ($0.01$ and $0.05$), and three seeds ($1$, $7$, and $42$).

Rather than repeating the full target pool, we select targets that cover different RHS cardinalities on each dataset. For Adult, the targets are $\text{education-num} \to \text{education}$ and $\text{education} \to \text{education-num}$. For Claims, we use $\text{Airport Name} \to \text{Airport Code}$, $\text{Airport Code} \to \text{Airport Name}$, $\{\text{Incident Date}, \text{Status}\} \to \text{Disposition}$. For Hospital, all three targets share Provider Number as the LHS, mapping to Emergency Service, State, and Hospital Name.

For a fair comparison, we also recompute the default strategies on exactly the same targets, error rates, and seeds used in the alternative-strategy batch. We refer to these results as the matched default results. The matched estimates therefore differ from the full-batch values reported earlier, which average over larger target sets. Table~\ref{tab:strategy-sensitivity} reports the matched results used in the comparisons below.

Most qualitative conclusions remain unchanged. Precision correlation remains positive for all three strength measures across all combinations of dataset, corruption mechanism, and replacement strategy. $\tau$ remains negatively correlated with recall in every combination. The broad dataset pattern for $\mu^+$ also persists, with recall correlation near zero or negative on Adult and Claims and positive on Hospital. One detail becomes clearer when the results are separated by corruption mechanism. On Adult under FD-aligned corruption, $\tau$ has a higher precision correlation than $g_3'$ with the matched default strategy, $0.7472$ versus $0.6599$. The same ordering holds under \texttt{WITHIN\_DOMAIN}, $0.7375$ versus $0.6577$, and under \texttt{FRESH\_VALUE}, $0.7781$ versus $0.6276$.

Under general corruption, the relative ordering depends on the replacement strategy. With the matched default \texttt{RANDOM\_COPY} strategy, $g_3'$ is higher than $\tau$, $0.7847$ versus $0.6236$. Under \texttt{WITHIN\_DOMAIN}, the two are nearly equal, with $0.7058$ for $g_3'$ and $0.7122$ for $\tau$. Under \texttt{FRESH\allowbreak\_VALUE}, $\tau$ is higher, $0.7763$ versus $0.6485$. Thus, replacement strategy changes the relative ordering of $\tau$ and $g_3'$ under general corruption, but not under FD-aligned corruption. Averaging over both corruption mechanisms in the full batch hides this difference.

\begin{table*}[t]
	\centering
	\caption{Precision correlation for $\tau$ and $g_3'$ on Adult, and recall correlation for $g_3'$ on Claims, split by corruption mechanism and replacement strategy, computed on the same matched subset of targets, error rates, and seeds. Each entry is the mean correlation across $12$ runs on Adult and $18$ runs on Claims. These matched estimates are not directly comparable to the full-batch defaults in Tables~\ref{tab:precision-correlation} and~\ref{tab:recall-correlation}, which aggregate $60$ runs on Adult and $165$ runs on Claims.}
	\label{tab:strategy-sensitivity}
	\begin{tabular}{llccc}
		\toprule
		Mechanism & Strategy & $\tau$ (Adult, precision) & $g_3'$ (Adult, precision) & $g_3'$ (Claims, recall) \\
		\midrule
		FD-aligned & Default (\texttt{CROSS\_CLASS}), matched & 0.7472 & 0.6599 & -0.4226 \\
		FD-aligned & \texttt{WITHIN\_DOMAIN} & 0.7375 & 0.6577 & -0.4223 \\
		FD-aligned & \texttt{FRESH\_VALUE} & 0.7781 & 0.6276 & -0.4484 \\
		General & Default (\texttt{RANDOM\_COPY}), matched & 0.6236 & 0.7847 & -0.3982 \\
		General & \texttt{WITHIN\_DOMAIN} & 0.7122 & 0.7058 & -0.4184 \\
		General & \texttt{FRESH\_VALUE} & 0.7763 & 0.6485 & -0.4449 \\
		\bottomrule
	\end{tabular}
\end{table*}

The negative association between $g_3'$ and recall on Claims is also robust to the replacement strategies tested here. Under the two alternative strategies, the $g_3'$--recall correlation remains negative under both corruption mechanisms, ranging from $-0.4484$ to $-0.4184$. The matched default values are $-0.4226$ under FD-aligned corruption and $-0.3982$ under general corruption. No sign reversal is observed on Claims under either alternative strategy.

\subsection{Search-Depth Sensitivity}
\label{sec:threats-depth}

The batch experiments in Section~\ref{sec:design} restrict the candidate space to $L\le3$. To test whether the findings depend on this bound, we extend the candidate space to $L\le4$ on the same target subset used in Section~\ref{sec:threats-strategy}. We retain the original replacement strategies, \texttt{CROSS\_CLASS} for FD-aligned corruption and \texttt{RANDOM\_COPY} for general corruption. We use both corruption mechanisms, two error rates ($0.01$ and $0.05$), and three seeds ($1$, $7$, and $42$), giving $96$ runs in total: $24$ on Adult, $36$ on Claims, and $36$ on Hospital. Within each run configuration, we compare the candidate set with $L\le3$ against the extended set with $L\le4$, while keeping all other parameters fixed.

The main qualitative conclusions remain unchanged when the candidate space is extended to $L\le4$. Precision correlation remains positive for all three strength measures on every dataset and under both corruption mechanisms at both search depths. For example, on Adult under FD-aligned corruption, the correlations at $L\le3$ are $0.8124$, $0.7472$, and $0.6599$ for $\mu^+$, $\tau$, and $g_3'$, respectively. Ranking mismatch also persists after adding $L=4$ candidates. Top-$k$ overlap remains close to zero for most measures and datasets, with no systematic improvement after the extension. Detection-cost regret also shows no systematic reduction. Some values decrease, while others increase substantially. For example, the regret of $g_3'$ rises from $1.83$ to $56.58$ on Adult under FD-aligned corruption and from $1.78$ to $56.44$ on Hospital under general corruption.

The association between $\rho_{ns}$ and recall is stronger after the candidate space is extended. Table~\ref{tab:depth-sensitivity} shows that the correlation increases in every dataset-mechanism combination. The additional $L=4$ candidates include more highly fragmented LHS partitions, yet the positive relationship between structural opportunity and recall remains and becomes stronger.

\begin{table}[t]
	\centering
	\caption{Spearman correlation between structural opportunity $\rho_{ns}$ and recall under $L\le3$ and $L\le4$, split by dataset and corruption mechanism.}
	\label{tab:depth-sensitivity}
	\begin{tabular}{llcc}
		\toprule
		Dataset & Mechanism & $L\le3$ & $L\le4$ \\
		\midrule
		Adult & FD-aligned & 0.7909 & 0.8823 \\
		Adult & General & 0.6668 & 0.7547 \\
		Hospital & FD-aligned & 0.9454 & 0.9616 \\
		Hospital & General & 0.9458 & 0.9605 \\
		Claims & FD-aligned & 0.9503 & 0.9670 \\
		Claims & General & 0.9429 & 0.9634 \\
		\bottomrule
	\end{tabular}
\end{table}

The same comparison also provides another check on the negative association between $g_3'$ and recall on Claims. With the original replacement strategies retained, the correlation remains negative at both search depths and becomes more negative after adding $L=4$ candidates. Under FD-aligned corruption, it changes from $-0.4226$ at $L\le3$ to $-0.5164$ at $L\le4$. Under general corruption, it changes from $-0.3982$ to $-0.4998$. Extending the candidate space therefore does not reverse the negative association under either corruption mechanism. The same negative sign also persists under the replacement-strategy changes examined in Section~\ref{sec:threats-strategy}.

\subsection{Sensitivity to Detection-Cost Weighting}
\label{sec:threats-weighting}

Section~\ref{sec:prelim-ldetect} uses $c_{FN}=c_{FP}=1$ as the default detection-cost setting. To test the sensitivity of the ranking results to this choice, we fix $c_{FP}=1$ and vary the relative false-negative cost $\lambda=c_{FN}/c_{FP}$ over $\{0.1,0.25,0.5,1,2,4,10\}$. Values below $1$ place greater relative weight on false positives, while values above $1$ place greater relative weight on false negatives. The tested range spans a factor of $100$ from the smallest to the largest cost ratio. These values are used as a sensitivity analysis rather than as estimates of application-specific costs.

On Hospital and Claims, $\mu^+$, $\tau$, and $g_3'$ select the same top candidate, as reported in Section~\ref{sec:ranking-utility}. Their regret is therefore identical at every tested value of $\lambda$. The selected candidate is never cost-optimal on either dataset, so the zero-regret fraction remains $0$ throughout the tested range.

On Adult, changing $\lambda$ also leaves the qualitative pattern unchanged. $\mu^+$ and $\tau$ never select a cost-optimal candidate at any tested value of $\lambda$. The candidate selected by $g_3'$ is cost-optimal in $51.67\%$ of runs, and this proportion remains unchanged across the full range of $\lambda$. Thus, even for $g_3'$, the strength-selected candidate is not cost-optimal in $48.33\%$ of runs.

The ranking mismatch therefore persists across all tested cost ratios. Changing the relative cost of false negatives and false positives affects the magnitude of detection loss, but it does not change the qualitative ranking results established in Section~\ref{sec:ranking-utility}. The equal-cost setting $c_{FN}=c_{FP}=1$ is therefore not responsible for the observed mismatch.

\subsection{Detection-Semantics Sensitivity}
\label{sec:threats-semantics}

The batch experiments use the \textsc{Non-Majority} detection semantics defined in Section~\ref{sec:design-detection}. To examine whether the observed relationships depend on this choice, we perform a small-scale paired comparison with the \textsc{Whole-Group} semantics on the same corrupted data. We use both Adult targets, one Hospital target, and one Claims target, with error rate $0.05$ and seeds $1$ and $42$. For each corrupted instance, we evaluate the same set of candidates under both detection semantics and compute the correlation of $\mu^+$ with precision and recall.

The precision relationship changes under \textsc{Whole-Group}. On Adult, the mean $\mu^+$--precision correlation decreases from $0.8311$ under \textsc{Non-Majority} to $0.6567$ under \textsc{Whole-Group}. The corresponding decrease is from $0.9470$ to $0.8573$ on Hospital and from $0.8033$ to $0.6657$ on Claims. The \textsc{Non-Majority} correlation is higher in every paired run in this comparison.

The difference is larger for recall. On Adult, the mean $\mu^+$--recall correlation changes from $-0.1044$ under \textsc{Non-Majority} to $-0.4721$ under \textsc{Whole-Group}. On Hospital, it decreases from $0.7919$ to $0.3446$. On Claims, it changes from $-0.1668$ to $-0.3461$. Thus, changing the detection semantics substantially changes the observed relationship between AFD strength and recall on all three datasets in this small-scale comparison.

These differences are consistent with the detector definitions. Under \textsc{Non-Majority}, only values below a unique majority are flagged, and a tied highest frequency produces no flags for the class. Under \textsc{Whole-Group}, every tuple in a conflicting class is flagged regardless of majority status. The two semantics therefore transform the same equivalence-class structure into different sets of flagged cells, which changes both precision and recall.

This comparison is intentionally limited in scale and does not support a full batch-level characterization of \textsc{Whole-Group}. It does show that the main conclusions should be interpreted within the \textsc{Non-Majority} semantics used throughout the batch experiments. Detection semantics affects the observed strength-value relationship and should therefore be treated as part of the scope of the results rather than as an interchangeable implementation choice.

\subsection{External Validity}
\label{sec:threats-external}

Our results are drawn from three real-world relations, Adult, Hospital, and Claims. These are established benchmark datasets from the RWD collection constructed by Parciak et al.~\cite{DBLP:journals/vldb/ParciakWHNPV25} for AFD measure evaluation, but our study still covers only three relations. These three relations do not cover the full range of structures that may occur in real data. Other datasets may differ in RHS cardinality, the balance between exact and approximate dependencies, the distribution of LHS equivalence classes, and the processes that generate both the data and the errors.

The main analysis restricts the candidate space to $L\le3$, and Section~\ref{sec:threats-depth} extends the search to $L\le4$ on a representative target subset. This does not establish robustness for substantially larger LHS sizes. The full target pools are not evaluated beyond $L=3$, and the analysis does not consider search depths beyond $L=4$.

As shown in Section~\ref{sec:threats-semantics}, the batch experiments use the \textsc{Non-Majority} detection semantics. A small-scale comparison with \textsc{Whole-Group} shows that changing the detection semantics can materially change the observed correlations. We therefore do not generalize the main batch-level conclusions to other detection semantics.

The sensitivity analyses support the robustness of the main conclusions to replacement strategies, search depths, and cost weightings tested here. Generalization to broader datasets, substantially deeper candidate spaces, or other detection semantics requires additional evidence.

\section{Related Work}
\label{sec:related}

\subsection{Approximate FD Strength Measures}
\label{sec:related-strength}

Measuring how closely a functional dependency holds on data that does not exactly satisfy it has a long history. Piatetsky-Shapiro and Matheus~\cite{3000767.3000784} are among the first to propose quantitative measures of dependency strength in large databases. Kivinen and Mannila~\cite{DBLP:journals/tcs/KivinenM95} formalize approximate inference of functional dependencies and introduce several error measures that remain influential. Giannella and Robertson~\cite{DBLP:journals/is/GiannellaR04} propose and compare a range of approximation measures based on these early definitions. Caruccio et al.~\cite{DBLP:journals/tkde/CaruccioDP16} provide a broader survey of relaxed functional dependencies, including variants beyond the strength measures considered here.

Among studies of AFD strength measures, the closest work to ours is Parciak et al.~\cite{DBLP:journals/vldb/ParciakWHNPV25}. They organize existing AFD strength measures into three families, \textsc{Logical}, \textsc{Violation}, and \textsc{Shannon}, and compare these measures on synthetic and real-world data. We adopt this taxonomy, using $\mu^+$ and $\tau$ from the \textsc{Logical} family and $g_3'$ from the \textsc{Violation} family. We exclude the \textsc{Shannon} family for the computational reasons discussed in Section~\ref{sec:datasets}. We also use the Adult, Hospital, and Claims relations from the RWD benchmark constructed by Parciak et al.

Prior work on AFD measures has primarily focused on how well a measure quantifies approximate dependency satisfaction and on its usefulness for discovering or ranking AFDs. The relationship between AFD strength and downstream error-detection performance has received much less attention. In particular, it remains unclear whether candidates with higher strength also provide higher precision or recall when used for error detection. Our study addresses this question directly.

\subsection{FD and AFD Discovery}
\label{sec:related-discovery}

A separate line of work studies how to discover FD or AFD candidates efficiently rather than how to evaluate a fixed candidate~\cite{DBLP:journals/tkde/LiuLLC12}. Huhtala et al.~\cite{DBLP:journals/cj/HuhtalaKPT99} introduce TANE for discovering exact and approximate functional dependencies. Papenbrock et al.~\cite{DBLP:journals/pvldb/PapenbrockEMNRZ15} compare seven FD discovery algorithms, while Bleifuss et al.~\cite{DBLP:conf/cikm/BleifussBFRW0PN16} use sampling to support approximate discovery on large datasets. More recent work addresses large-scale discovery~\cite{DBLP:journals/tkde/WanHWL24}, relaxed dependencies based on multi-attribute dominance~\cite{DBLP:journals/tkde/CaruccioDNP21}, hitting-set-based FD discovery~\cite{DBLP:journals/pacmmod/BleifussPBSN24}, and top-$k$ discovery that reduces redundancy among returned dependencies~\cite{DBLP:journals/corr/abs-2601-10130}.

A closely related discovery setting is considered by Wan and Han~\cite{DBLP:journals/corr/abs-2605-24925}, who discover top-$k$ AFDs by ranking candidates with $\mu^+$ and pruning low-strength branches during search. Mandros et al.~\cite{DBLP:conf/kdd/MandrosBV17} similarly combine a reliability-oriented dependency measure with an algorithm for discovering high-scoring dependencies. As discussed in Section~\ref{sec:datasets}, we exclude the \textsc{Shannon} family used by Mandros et al. from our comparison for computational reasons.

Our experiments deliberately separate candidate discovery from candidate evaluation. We enumerate the full candidate space up to a fixed LHS bound without strength-based pruning, so the candidate pool is not biased toward dependencies favored by a particular discovery algorithm. We assume that the candidate space is available and ask whether ranking candidates by AFD strength tracks their error-detection value. Improving discovery efficiency does not resolve this evaluation question.

\subsection{Constraint-Based Error Detection}
\label{sec:related-cleaning}

Dependencies and other integrity constraints are widely used to detect or repair inconsistent data~\cite{DBLP:conf/icde/ChuIP13,DBLP:conf/sigmod/DallachiesaEEEIOT13,DBLP:journals/tods/FanGJK08,DBLP:journals/pvldb/KaminskyPN24}. Ilyas and Chu~\cite{DBLP:books/acm/IlyasC19} and Abedjan et al.~\cite{DBLP:series/synthesis/2018Abedjan} provide broader overviews of constraint-based data cleaning and data profiling. Existing work shows how constraints can be used to flag or repair inconsistent data once the constraints are available. Our focus is how to choose among available approximate dependencies. Given several candidates with different strength scores, we ask whether these scores identify which candidates are more useful for error detection. This question arises before any cleaning or repair is performed.

Zhang et al.~\cite{DBLP:conf/sigmod/ZhangGR20} study how noise affects the reliable discovery of functional dependencies. Our setting differs from theirs. We take the candidate dependency as given and ask whether its strength score, computed on a particular data instance, is associated with its error-detection performance on that instance. Their work studies how noise affects dependency discovery, whereas ours studies whether dependency strength is informative for detecting errors.

\subsection{The Missing Link: Strength Versus Task Utility}
\label{sec:related-gap}

Systematic assessment of data quality remains underexplored in the broader literature. Much existing work focuses instead on tools and techniques for improving data quality. Mohammed et al.~\cite{DBLP:journals/sigmod/MohammedEHNS25} note that, despite the availability of many data-cleaning tools, systematic methods for assessing data quality across different dimensions remain limited. This makes it difficult to determine whether a cleaning effort has actually improved data quality.

Our study addresses a more specific version of this problem. AFD strength measures evaluate how well a score reflects approximate dependency satisfaction, but they do not directly establish whether a higher score also means better error-detection performance. Dependency discovery methods focus on finding candidates efficiently, leaving open which discovered candidate is most useful for error detection. Constraint-based cleaning shows how dependencies can be used to detect and repair errors, but usually assumes that the dependency has already been selected. It therefore does not address whether strength provides a reliable basis for choosing among several available candidates. Recent work has also questioned whether conventional rule-validity criteria are sufficient indicators of rule quality. Martin et al.~\cite{DBLP:journals/pvldb/MartinARQ25} show that existing denial-constraint discovery methods can return large numbers of false constraints under the current DC validity definition. Their study concerns the validity of discovered denial constraints, whereas we examine whether AFD strength reflects downstream error-detection value.

Recent work on top-$k$ rule discovery has recognized that objective rule measures alone may be insufficient to capture usefulness. Fan et al.~\cite{DBLP:journals/pacmmod/FanHWX23} combine objective measures such as support and confidence with a subjective measure learned from user feedback. Fan et al.~\cite{DBLP:journals/pacmmod/FanHXZ24} further consider relevance and diversity and evaluate discovered top-$k$ rules using error-detection outcomes. Han et al.~\cite{DBLP:journals/tkde/HanCHMQ26} learn user-guided embeddings for relevant and diversified top-$k$ rule discovery while reducing interaction and computational cost.

These approaches improve rule selection by adding learned or user-supplied signals. We ask how much error-detection utility is already captured by established AFD strength measures before such signals are added. To the best of our knowledge, prior work has not systematically tested whether AFD strength itself tracks the downstream error-detection utility of individual dependencies. Our study addresses this gap by evaluating AFD strength directly against error-detection outcomes and candidate-selection performance.

\section{Conclusion}
\label{sec:conclusion}

This paper examines whether higher AFD strength also implies greater value for error detection. Across three real-world relations, $1{,}155$ batch runs, and $513{,}480$ candidate evaluations, we find that strength is positively associated with precision but does not track recall consistently. Strength-based rankings often differ from those based on actual detection performance. An important part of the recall mismatch is associated with structural detection opportunity: recall depends on whether the LHS partition contains enough non-singleton structure for relevant errors to produce observable conflicts, whereas conventional strength measures do not directly encode this property.

This work makes three contributions. First, we define $V_{use}$ and $L_{detect}$ to evaluate error-detection value separately from approximate dependency satisfaction. Second, we provide a systematic empirical comparison of $\mu^+$, $\tau$, and $g_3'$ under multiple corruption mechanisms, replacement strategies, search depths, and cost settings, together with a small-scale paired comparison of detection semantics. Third, we identify structural detection opportunity as a measurable factor that is much more closely associated with recall than the tested strength measures, while also showing that corruption mechanism, RHS cardinality, how often the RHS attribute appears in other LHSs, and LHS uniqueness do not fully explain the remaining variation.

The main conclusion is that AFD strength and error-detection value are related but different. Although a higher strength score can indicate better precision, it does not reliably identify the dependencies with better recall or lower detection cost. Strength should therefore be evaluated for the downstream task in which it is used rather than assumed to represent task value.

\bibliographystyle{spmpsci}      
\bibliography{references}   

\begin{thebibliography}{10}
\providecommand{\url}[1]{{#1}}
\providecommand{\urlprefix}{URL }
\expandafter\ifx\csname urlstyle\endcsname\relax
  \providecommand{\doi}[1]{DOI~\discretionary{}{}{}#1}\else
  \providecommand{\doi}{DOI~\discretionary{}{}{}\begingroup
  \urlstyle{rm}\Url}\fi

\bibitem{DBLP:series/synthesis/2018Abedjan}
Abedjan, Z., Golab, L., Naumann, F., Papenbrock, T.: Data Profiling.
\newblock Synthesis Lectures on Data Management. Morgan {\&} Claypool
  Publishers (2018)

\bibitem{DBLP:conf/cikm/BleifussBFRW0PN16}
Bleifu{\ss}, T., B{\"{u}}low, S., Frohnhofen, J., Risch, J., Wiese, G., Kruse,
  S., Papenbrock, T., Naumann, F.: Approximate discovery of functional
  dependencies for large datasets.
\newblock In: Proceedings of the 25th {ACM} International Conference on
  Information and Knowledge Management, {CIKM} 2016, pp. 1803--1812. {ACM}
  (2016)

\bibitem{DBLP:journals/pacmmod/BleifussPBSN24}
Bleifu{\ss}, T., Papenbrock, T., Bl{\"{a}}sius, T., Schirneck, M., Naumann, F.:
  Discovering functional dependencies through hitting set enumeration.
\newblock Proc. {ACM} Manag. Data \textbf{2}(1), 43:1--43:24 (2024)

\bibitem{DBLP:conf/sigmod/BohannonFFR05}
Bohannon, P., Flaster, M., Fan, W., Rastogi, R.: A cost-based model and
  effective heuristic for repairing constraints by value modification.
\newblock In: F.~{\"{O}}zcan (ed.) Proceedings of the {ACM} {SIGMOD}
  International Conference on Management of Data, pp. 143--154. {ACM} (2005)

\bibitem{DBLP:journals/tkde/CaruccioDNP21}
Caruccio, L., Deufemia, V., Naumann, F., Polese, G.: Discovering relaxed
  functional dependencies based on multi-attribute dominance.
\newblock {IEEE} Trans. Knowl. Data Eng. \textbf{33}(9), 3212--3228 (2021)

\bibitem{DBLP:journals/tkde/CaruccioDP16}
Caruccio, L., Deufemia, V., Polese, G.: Relaxed functional dependencies - {A}
  survey of approaches.
\newblock {IEEE} Trans. Knowl. Data Eng. \textbf{28}(1), 147--165 (2016)

\bibitem{DBLP:conf/icde/ChuIP13}
Chu, X., Ilyas, I.F., Papotti, P.: Holistic data cleaning: Putting violations
  into context.
\newblock In: 29th {IEEE} International Conference on Data Engineering, {ICDE}
  2013, Brisbane, Australia, April 8-12, 2013, pp. 458--469. {IEEE} Computer
  Society (2013)

\bibitem{DBLP:conf/sigmod/DallachiesaEEEIOT13}
Dallachiesa, M., Ebaid, A., Eldawy, A., Elmagarmid, A.K., Ilyas, I.F., Ouzzani,
  M., Tang, N.: {NADEEF:} a commodity data cleaning system.
\newblock In: K.A. Ross, D.~Srivastava, D.~Papadias (eds.) Proceedings of the
  {ACM} {SIGMOD} International Conference on Management of Data, {SIGMOD} 2013,
  pp. 541--552. {ACM} (2013)

\bibitem{DBLP:journals/tods/FanGJK08}
Fan, W., Geerts, F., Jia, X., Kementsietsidis, A.: Conditional functional
  dependencies for capturing data inconsistencies.
\newblock {ACM} Trans. Database Syst. \textbf{33}(2), 6:1--6:48 (2008)

\bibitem{DBLP:journals/pacmmod/FanHWX23}
Fan, W., Han, Z., Wang, Y., Xie, M.: Discovering top-k rules using subjective
  and objective criteria.
\newblock Proc. {ACM} Manag. Data \textbf{1}(1), 70:1--70:29 (2023)

\bibitem{DBLP:journals/pacmmod/FanHXZ24}
Fan, W., Han, Z., Xie, M., Zhang, G.: Discovering top-k relevant and
  diversified rules.
\newblock Proc. {ACM} Manag. Data \textbf{2}(4), 195:1--195:28 (2024)

\bibitem{DBLP:journals/is/GiannellaR04}
Giannella, C., Robertson, E.L.: On approximation measures for functional
  dependencies.
\newblock Inf. Syst. \textbf{29}(6), 483--507 (2004)

\bibitem{DBLP:journals/tkde/HanCHMQ26}
Han, Z., Chen, W., Han, Y., Mao, R., Qin, J.: Fast diversified top-k rule
  discovery via user-guided embeddings.
\newblock {IEEE} Trans. Knowl. Data Eng. \textbf{38}(3), 1739--1753 (2026)

\bibitem{hollander2014nonparametric}
Hollander, M., Wolfe, D.A., Chicken, E.: Nonparametric Statistical Methods, 3rd
  edn.
\newblock John Wiley \& Sons, Hoboken, NJ (2014).
\newblock \doi{10.1002/9781119196037}

\bibitem{DBLP:journals/cj/HuhtalaKPT99}
Huhtala, Y., K{\"{a}}rkk{\"{a}}inen, J., Porkka, P., Toivonen, H.: {TANE:} an
  efficient algorithm for discovering functional and approximate dependencies.
\newblock Comput. J. \textbf{42}(2), 100--111 (1999)

\bibitem{DBLP:books/acm/IlyasC19}
Ilyas, I.F., Chu, X.: Data Cleaning, \emph{{ACM} Books}, vol.~28.
\newblock {ACM} (2019)

\bibitem{DBLP:journals/pvldb/KaminskyPN24}
Kaminsky, Y., Pena, E.H.M., Naumann, F.: Incremental detection of denial
  constraint violations.
\newblock Proc. {VLDB} Endow. \textbf{18}(4), 1000--1012 (2024)

\bibitem{DBLP:journals/tcs/KivinenM95}
Kivinen, J., Mannila, H.: Approximate inference of functional dependencies from
  relations.
\newblock Theor. Comput. Sci. \textbf{149}(1), 129--149 (1995)

\bibitem{DBLP:journals/tkde/LiuLLC12}
Liu, J., Li, J., Liu, C., Chen, Y.: Discover dependencies from data - {A}
  review.
\newblock {IEEE} Trans. Knowl. Data Eng. \textbf{24}(2), 251--264 (2012)

\bibitem{DBLP:conf/kdd/MandrosBV17}
Mandros, P., Boley, M., Vreeken, J.: Discovering reliable approximate
  functional dependencies.
\newblock In: Proceedings of the 23rd {ACM} {SIGKDD} International Conference
  on Knowledge Discovery and Data Mining, pp. 355--363. {ACM} (2017)

\bibitem{DBLP:journals/pvldb/MartinARQ25}
Martin, A., de~Almeida, E.C., Romero, O., Queralt, A.: How and why false denial
  constraints are discovered.
\newblock Proc. {VLDB} Endow. \textbf{18}(10), 3477--3489 (2025)

\bibitem{DBLP:journals/sigmod/MohammedEHNS25}
Mohammed, S., Ehrlinger, L., Harmouch, H., Naumann, F., Srivastava, D.: The
  five facets of data quality assessment.
\newblock {SIGMOD} Rec. \textbf{54}(2), 18--27 (2025)

\bibitem{DBLP:journals/pvldb/PapenbrockEMNRZ15}
Papenbrock, T., Ehrlich, J., Marten, J., Neubert, T., Rudolph, J.,
  Sch{\"{o}}nberg, M., Zwiener, J., Naumann, F.: Functional dependency
  discovery: An experimental evaluation of seven algorithms.
\newblock Proc. {VLDB} Endow. \textbf{8}(10), 1082--1093 (2015)

\bibitem{DBLP:journals/vldb/ParciakWHNPV25}
Parciak, M., Weytjens, S., Hens, N., Neven, F., Peeters, L.M., Vansummeren, S.:
  Measuring approximate functional dependencies: a comparative study.
\newblock {VLDB} J. \textbf{34}(4), 56 (2025)

\bibitem{parciak_2023_8098909}
Parciak, M., Weytjens, S., Neven, F., Hens, N., Peeters, L.M., Vansummeren, S.:
  Annotated benchmark of real-world data for approximate functional dependency
  discovery (2023).
\newblock \doi{10.5281/zenodo.8098909}.
\newblock \urlprefix\url{https://doi.org/10.5281/zenodo.8098909}

\bibitem{3000767.3000784}
Piatetsky-Shapiro, G., Matheus, C.J.: Measuring data dependencies in large
  databases.
\newblock In: Proceedings of the 2nd International Conference on Knowledge
  Discovery in Databases, AAAIWS'93, p. 162–173. AAAI Press (1993)

\bibitem{DBLP:journals/corr/abs-2607-22219}
Wan, X., Han, X.: Not every dependency is worth discovering: Toward
  value-driven data dependency discovery.
\newblock CoRR \textbf{abs/2607.22219} (2026)

\bibitem{DBLP:journals/corr/abs-2601-10130}
Wan, X., Han, X.: Redundancy-driven top-k functional dependency discovery.
\newblock CoRR \textbf{abs/2601.10130} (2026)

\bibitem{DBLP:journals/corr/abs-2605-24925}
Wan, X., Han, X.: Top-k approximate functional dependency discovery.
\newblock CoRR \textbf{abs/2605.24925} (2026)

\bibitem{DBLP:journals/tkde/WanHWL24}
Wan, X., Han, X., Wang, J., Li, J.: Efficient discovery of functional
  dependencies on massive data.
\newblock {IEEE} Trans. Knowl. Data Eng. \textbf{36}(1), 107--121 (2024)

\bibitem{DBLP:conf/sigmod/ZhangGR20}
Zhang, Y., Guo, Z., Rekatsinas, T.: A statistical perspective on discovering
  functional dependencies in noisy data.
\newblock In: Proceedings of the 2020 International Conference on Management of
  Data, pp. 861--876. {ACM} (2020)

\end{thebibliography}


\end{document}